\documentclass[useAMS,usenatbib,babel]{mn2e}

\usepackage[english,english]{babel}
\usepackage{amsmath}
\usepackage{amssymb,amsfonts,textcomp}
\usepackage{array}
\usepackage{supertabular}
\usepackage{hhline}
\usepackage{hyperref}
\usepackage[usenames]{color}
\hypersetup{dvips, colorlinks=true, linkcolor=black, citecolor=black, filecolor=blue, urlcolor=blue}
\usepackage[dvips]{graphicx}

\def\gtrsim{\lower.5ex\hbox{$\; \buildrel > \over \sim \;$}}
\usepackage{graphicx}

\newcommand{\nofeed}{\mbox{\emph{nofeedback}}}
\newcommand{\noAGN}{\mbox{\emph{noAGN}}}
\newcommand{\AGN}{\mbox{\emph{AGN}}}

\definecolor{grey}{rgb}{0.75,0.75,0.75}
\definecolor{Orange}{rgb}{1.0,0.5,0.15}
\definecolor{brown}{rgb}{0.7,0.25,0.0}
\definecolor{pink}{rgb}{1.0,0.5,0.5}
\definecolor{darkerred}{rgb}{0.8,0,0}
\definecolor{darkerblue}{rgb}{0,0,0.8}
\definecolor{Blue}{rgb}{0,0.08,0.65}
\definecolor{Red}{rgb}{0.65,0.08,0.05}
\definecolor{Green}{rgb}{0.15,0.45,0.25}

\begin{document}

\author[Y. Dubois et al. ]{
\parbox[t]{\textwidth}{
Yohan Dubois$^{1,2}$\thanks{E-mail: dubois@iap.fr}, Christophe Pichon$^{1,3}$, Julien Devriendt$^{2,4}$, Joseph Silk$^{1,2,5}$,\\ Martin Haehnelt$^{6}$, Taysun  Kimm$^{2}$ and Adrianne~Slyz$^{2}$ }
\vspace*{6pt} \\
$^{1}$ Institut d'Astrophysique de Paris, UMR 7095, CNRS, UPMC Univ. Paris VI, 98 bis boulevard Arago, 75014 Paris, France\\
$^{2}$ Sub-department of Astrophysics, University of Oxford, Keble Road, Oxford OX1 3RH, United Kingdom\\
$^{3}$ CEA Saclay, DSM/IPhT, B\^atiment 774, 91191 Gif-sur-Yvette, France\\
$^{4}$ Observatoire de Lyon, UMR 5574, 9 avenue Charles Andr\'e, Saint Genis Laval 69561, France\\
$^{5}$ Department of Physics and Astronomy, The Johns Hopkins University Homewood Campus, Baltimore MD 21218, USA\\
$^{6}$ Institute of Astronomy and Kavli Institute for Cosmology, Madingley Road, Cambridge CB3 0HA, United Kingdom\\
}
\date{Accepted . Received ; in original form }

\title[AGN and high-redshift galaxy formation]
{Blowing cold flows away: the impact of early AGN activity on the formation of a brightest cluster galaxy progenitor}

\maketitle

\begin{abstract}
{Supermassive black holes (BH) are powerful sources of energy that are already in place at very early epochs of the Universe (by $z=6$). Using hydrodynamical simulations of the formation of a massive $M_{\rm vir}=5\times 10^{11}\, \rm M_\odot$ halo by $z=6$ (the most massive progenitor of a cluster of $M_{\rm vir}=2\times 10^{15}\, \rm M_\odot$ at $z=0$), we  evaluate the impact of Active Galactic Nuclei (AGN) on galaxy mass content, BH self-regulation, and gas distribution inside this massive halo. We find that SN feedback has a marginal influence on the stellar structure, and no influence on the mass distribution on large scales. In contrast, AGN feedback alone is able to significantly alter the stellar-bulge mass content by quenching  star formation  when the BH is self-regulating, and by depleting the cold gas reservoir in the centre of the galaxy. The growth of the BH proceeds first by a rapid Eddington-limited period fed by direct cold filamentary infall. When the energy delivered by the AGN is sufficiently large to unbind the cold gas of the bulge, the accretion of gas onto the BH is maintained both by  smooth gas inflow and clump migration through the galactic disc triggered by merger-induced  torques. The feedback from the AGN has also a severe consequence on the baryon mass content within the halo, producing large-scale hot superwinds, able to blow away some of the cold filamentary material from the centre and reduce the baryon fraction by more than 30 per cent within the halo's virial radius. Thus in the very young universe, AGN feedback is likely to be a key process, shaping the properties of the most massive galaxies.}
\end{abstract}


\section{Introduction}

Supermassive black holes (BH) are ubiquitous in the centres of galaxies with masses scaling with those of their hosts~\citep{magorrianetal98, tremaineetal02, haring&rix04}, and with the most massive BHs found in the centres of the brightest cluster galaxies (BCGs).
These observations suggest that a self-regulating process is at play in galaxy interiors involving some source of feedback energy produced by accretion of material onto their central BHs~\citep{silk&rees98, king03, wyithe&loeb03}.
The reality of such Active Galactic Nuclei (AGN) is confirmed by direct detections of radio jets and X-ray cavities in the core of groups and clusters of galaxies~\citep[e.g.][]{boehringeretal93}, and also by observations of broad absorption lines in the spectra of quasars~\citep[e.g.][]{chartasetal03}

Semi-analytical models of galaxy formation~\citep{crotonetal06, boweretal06, cattaneoetal06, somervilleetal08} with AGN feedback quench star formation in massive elliptical galaxies, and reproduce the bright end of the galaxy luminosity function with a sharp cut-off that is in good agreement with observations~\citep{belletal03, panteretal07}.
Self-regulated growth of BHs with AGN feedback has also been modeled in various hydrodynamical cosmological simulations, reproducing the tight constraints on BH and galaxy relationships at $z=0$, and confirming that this feedback strongly affects massive galaxy formation~\citep{sijackietal07, dimatteoetal08, booth&schaye09, duboisetal12agnmodel, kimmetal12}.
These numerical models manage to prevent the cooling catastrophe by regulating cooling flows in galaxy clusters, a necessary condition for obtaining massive galaxies with low star formation rates~\citep{duboisetal10, duboisetal11, mccarthyetal10, mccarthyetal11, teyssieretal11}.

For scenarios without strong feedback, density profiles show large mass concentrations in the core regions of galaxy clusters that are at odds with observations.
This is a consequence of the catastrophic cooling of the  gas, which drives too many baryons into the cores of halos and consequently increases their gravitational potential wells via adiabatic contraction~\citep{blumenthaletal86, gnedinetal04}, steepening their DM density profiles.
This generic process has been suggested to occur early on in the mass assembly of clusters, and could be avoided by preventing baryons from cooling excessively and/or by blowing out large quantities of gas~\citep{read&gilmore05, pontzen&governato12, martizzietal12b}.
As the gas in the intra-cluster medium is approximately in local hydrostatic equilibrium, gas thermodynamical properties are likely to be set up by the gas density distribution and the degree to which adiabatic contraction has  developed or has been prevented.
Therefore without AGN feedback, gas in the intra-cluster medium is strongly depleted, giving unrealistic values for the temperature and entropy in cluster cores compared to X-ray observations~\citep{tornatoreetal03, valdarnini03, nagaietal07}.
The feedback activity at high redshift is likely to play an essential role in establishing and shaping some of the properties in clusters of galaxies, including the amount of gas present within the halo, the quantity of cold gas available to form stars, and the total stellar mass.

Indeed, observations of distant and bright quasars suggest that strong feedback activity was present in the early epochs of galaxy formation before $z=6$~\citep{fanetal06,jiangetal09,mortlocketal11}.
Massive BHs with several $10^{8-9} \, \rm M_\odot$ are already in place, deeply embedded in dense gaseous regions, with rapid accretion onto them at rates close to the Eddington limit~\citep{volonteri&rees06, willottetal10, treisteretal10} triggering powerful outflows.
But, it is still very difficult to determine from high redshift observations the impact of early AGN feedback on the gas content of galaxies and the mass distribution within their halos.

There are fundamental differences in the nature of the gas accreted onto galaxies at different epochs: red and massive galaxies are surrounded by hot and diffuse gas such as the intra-cluster medium detected in X-rays~\citep{gurskyetal71, formanetal72, markevitchetal98, vikhlininetal05, prattetal07}, while theoretical and numerical models of $\Lambda$CDM universes predict that young and gas-rich galaxies obtain their gas through cold isothermal and collimated flows~\citep{binney77, birnboim&dekel03, keresetal05, keresetal09, ocvirketal08, brooksetal09, dekeletal09} that are still to be confirmed by direct detection~\citep{kimmetal11cfdet, goerdtetal12}.
Though the interaction of AGN (radio jets) with their surrounding, hot intra-cluster gas has been extensively studied~\citep{ommaetal04, ruszkowskietal04, bruggenetal05, brighenti&mathews06, sternbergetal07, gasparietal11}, little is known about how quasars at high redshift interact with cold infalling streams of gas, and in particular whether these collimated structures can survive the energy released through these powerful outflows.
While feedback from supernova explosions alone is not able to stop the streaming of cold flows towards galaxies~\citep{powelletal11}, the larger energies involved in AGN activity can potentially alter the motions and the morphology of the filamentary gas.
In a large cosmological simulation box,~\cite{dimatteoetal12} (see also~\citealp{khandaietal12}) have shown that the primordial, most massive BHs can grow uninterrupted via cold filamentary infall, and despite AGN feedback, can acquire up to several billion solar masses at $z=6$ in the most massive halos ($M_{\rm vir}\simeq10^{12-13}\, \rm M_\odot$) and inside the most active galaxies at this redshift, consistent with the observations of the most distant quasars~\citep{wangetal10, wangetal11}.
Similar conclusions have been reached previously by other authors using resimulations of individual massive objects~\citep{lietal07, sijackietal09}.

Using well-resolved hydrodynamical cosmological simulations, \cite{duboisetal12angmom} have shown that, in the absence of feedback, the accretion onto the central bulge (and potentially a central BH) of massive galaxies is two-fold: during an early phase, cold streams feed directly the centre of the galaxy; in a second phase, they contribute to the formation of proto-galactic discs by gaining angular momentum~\citep{pichonetal11}.
A large amount of gas is accreted onto the disc, maintains it to be gravitationally unstable and will rapidly feed the compact bulge through migration of clumps~\citep[see also][]{bournaudetal11}.

We will address here the question of massive galaxy formation at high redshift from another angle while focusing on the impact of AGN feedback onto the early assembly phase of a progenitor of a massive cluster of galaxies - but not necessarily the most massive object at $z=6$ - using sufficient resolution to capture the detailed structure of the galaxy and of the inflow/outflow onto it ($\Delta x=15$~pc).
We will question whether feedback from AGN is able to rapidly quench  star formation inside the central galaxy, and what kind of outflows can form from the central source of energy.
Our model of AGN feedback has been calibrated to reproduce observational constraints at $z=0$ (such as BH mass density and $M_{\rm BH}$-$M_{\rm b}$, \citealp{duboisetal12agnmodel})  using kpc-resolution hydrodynamical cosmological simulations, and no recalibration of the model is performed in what follows, unless stated explicitly.
Standard recipes are used to describe the physics of star formation and stellar feedback.
Hence in this paper we investigate the consequences of such a ``canonical'' feedback model on the early galaxy evolution and its associated inflow/outflow.

The paper is organised as follows: in section~\ref{section:numerics}, we present the details of the numerical implementation. We report the impact of AGN feedback on the galaxy properties (section~\ref{section:bulge}), the accretion rate onto the central BH (section~\ref{section:energy}), and the impact on the gas content within the halo (section~\ref{section:outflow}). Finally we discuss the  implications of such strong AGN feedback (section~\ref{section:conclusion}).

\section{Numerical set-up}
\label{section:numerics}

The initial conditions are designed to yield a rare and massive cluster halo with a virial mass of $M_{\rm vir}(z=0)=2 \times 10^{15}\, \rm M_\odot$ at $z=0$.
We assume a $\Lambda$CDM cosmology with total matter density $\Omega_{m}=0.27$, baryon density $\Omega_b=0.045$, dark energy density $\Omega_{\Lambda}=0.73$, amplitude of the matter power spectrum $\sigma_8=0.8$ and  Hubble constant $H_0=70\, \rm km\, s^{-1} \, \rm Mpc^{-1}$ consistent with the WMAP 7-year data \citep{komatsuetal11}.
The box size of our simulations is $L_{\rm box}=100\,  h^{-1}\, \rm Mpc$, within which we define a high resolution region where high resolution DM particles ($M_{\rm res}=9\times 10^{5}\, h^{-1}\, \rm M_{\odot}$) fill a volume of size $2\, r_{\rm vir}$ around the main progenitor of the targeted massive cluster halo at $z=6$.
The mass of DM particles is slowly downgraded from inside-out.
The mass high resolution region fills a sphere of  $2.3\,h^{-1}\, \rm Mpc$ comoving radius in the initial conditions, then lower resolution DM particles fill concentric shells as follow: $7\times10^6\, h^{-1}\,\rm M_{\odot}$ from $2.3$-$2.8\, h^{-1}\,\rm Mpc$, $6\times10^7\, h^{-1}\,\rm M_{\odot}$ from $2.8$-$3.5\, h^{-1}\,\rm Mpc$, $5\times10^8\, h^{-1}\,\rm M_{\odot}$ from $3.5$-$5\, h^{-1}\,\rm Mpc$, $4\times10^9\, h^{-1}\,\rm M_{\odot}$ from $5$-$7.5\, h^{-1}\,\rm Mpc$, $3\times10^{10}\, h^{-1}\,\rm M_{\odot}$ from $7.5$-$15\, h^{-1}\,\rm Mpc$, and $2\times10^{11}\, h^{-1}\,\rm M_{\odot}$ filling the rest of the box.
To smooth out the contribution of DM particles to the gravitational potential in highly refined regions of the gas flows, their mass is projected up to the level 16 of the grid (while the grid reaches up level 20 at $z=6$) for the computation of the gravitational potential that correspond to an effective softening length of $300$~pc at redshift 6.
This is 20 times the softening length for the hydro solver at this redshift.
The progenitor of the halo has a mass and virial radius at $z=6$ of $M_{\rm vir}(z=6)=5 \times 10^{11}\, \rm M_\odot$ and $r_{\rm vir}=35$~kpc.

The simulations are run with the Adaptive Mesh Refinement code {\sc ramses} \citep{teyssier02}.
The evolution of the gas is followed using a second-order unsplit Godunov scheme for the Euler equations.
The HLLC Riemann solver with a first-order MinMod Total Variation Diminishing scheme to reconstruct the interpolated variables from their cell-centered values is used to compute fluxes at cell interfaces.
Collisionless particles (DM and star particles) are evolved using a particle-mesh solver with a Cloud-In-Cell interpolation.
The mesh is refined up to $\Delta x=15$~pc using a quasi-Lagrangian strategy: when more than 8 DM particles lie in a cell, or if the baryon density is larger than 8 times the initial DM resolution.

Gas is allowed to cool by H and He cooling with a contribution from metals using a~\cite{sutherland&dopita93} model for temperatures above $10^4$~K, and metal fine-structure cooling below $10^4$~K following~\cite{rosen&bregman95}.
The gas temperature is not allowed  to fall below $T_0=100$~K.
Heating from a uniform UV background takes place after redshift $z_{\rm reion}=8.5$ following~\cite{haardt&madau96}.
Metallicity is modeled as a passive variable for the gas (whose composition is assumed to be solar) and is altered by the injection of gas ejecta during supernovae (SNe) explosions, assuming a constant yield of 0.1.
We assume an initial metallicity of $Z=10^{-3}\, \rm Z_\odot$.

The star formation process is modeled with a Schmidt law:
$\dot \rho_*= \epsilon_* {\rho / t_{\rm ff}}\, ,$ where $\dot \rho_*$ is the star formation rate density, $\epsilon_*$ the constant star formation efficiency, and $t_{\rm ff}$ the local free-fall time of the gas.
We choose a low star formation efficiency $\epsilon_*=0.01$ consistent with observations of giant molecular clouds~\citep{krumholz&tan07}.
Star formation is allowed in regions exceeding a gas density threshold of $\rho_0=50\, \rm H\, cm^{-3}$.
The gas pressure is artificially enhanced above $\rho > \rho_0$ assuming a polytropic equation of state $T=T_0(\rho/\rho_0)^{\kappa-1}$ with polytropic index $\kappa=2$ to avoid excessive gas fragmentation.
Feedback from SNe releases $10^{51}$~erg per 10~$\rm M_\odot$ assuming a Salpeter IMF where $\eta_{\rm SN}=0.1$ fraction of the IMF is composed of type II core-collapse SNe.
The energy from SNe is coupled to the gas with a kinetic implementation from~\cite{dubois&teyssier08winds}, where mass, momentum and kinetic energy profiles are imposed to mimic a Sedov blast wave.
Neither feedback from stellar winds nor type Ia SNe are taken into account.

\begin{figure*}
  \centering{\resizebox*{!}{5.2cm}{\includegraphics{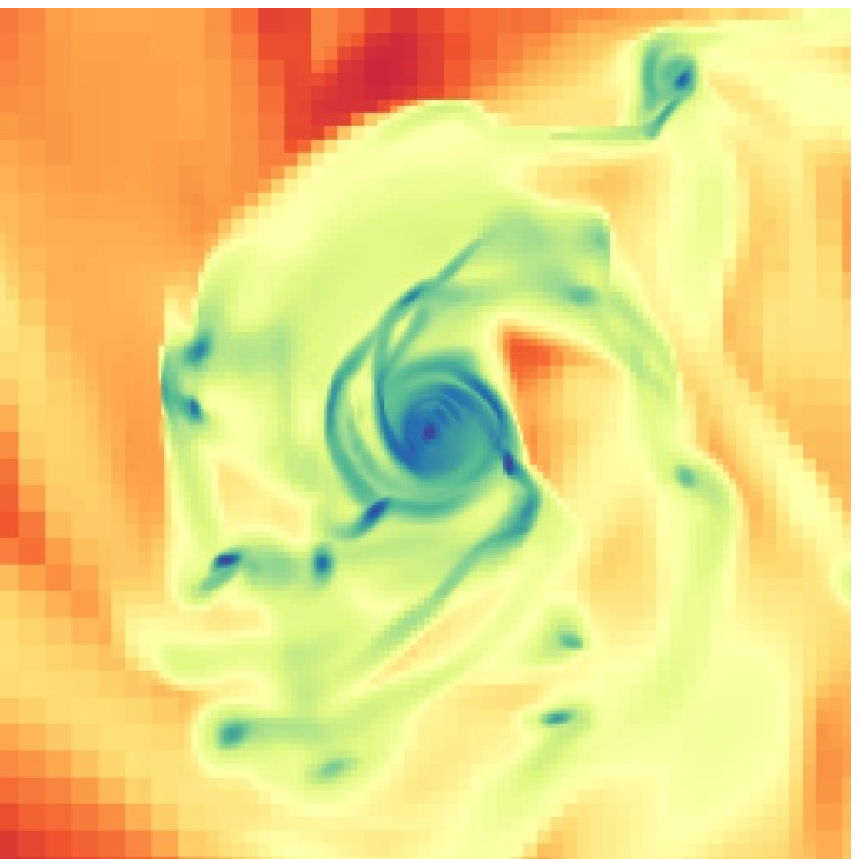}}}
  \centering{\resizebox*{!}{5.2cm}{\includegraphics{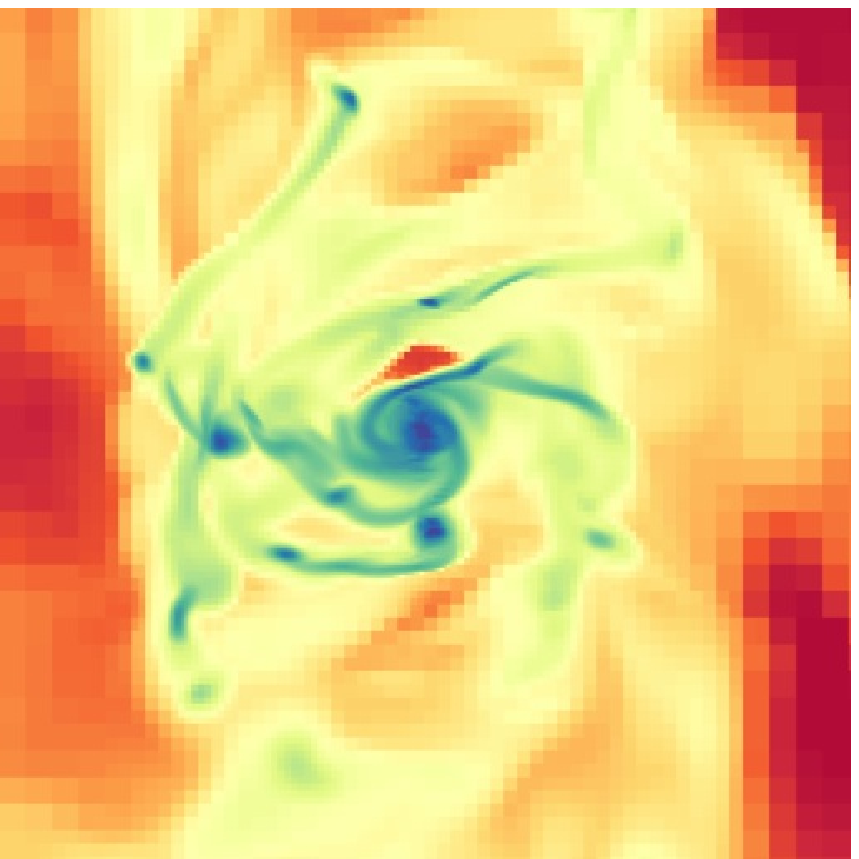}}}
  \centering{\resizebox*{!}{5.2cm}{\includegraphics{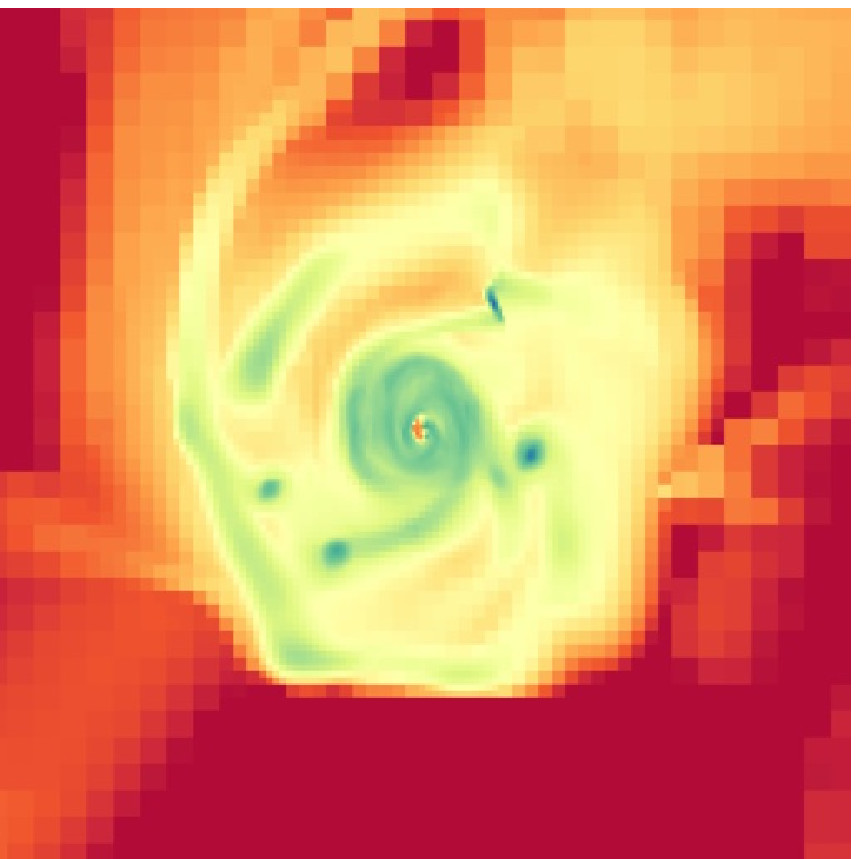}}}\hspace{0.2cm}
  \centering{\resizebox*{!}{5.2cm}{\includegraphics{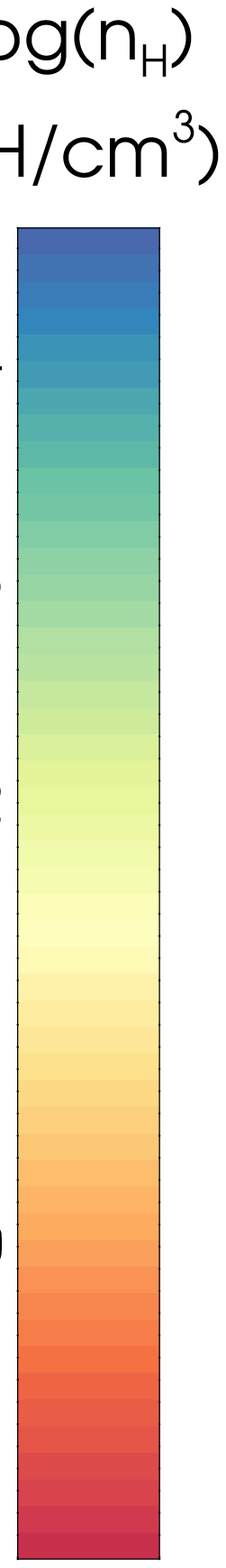}}}
  \centering{\resizebox*{!}{5.2cm}{\includegraphics{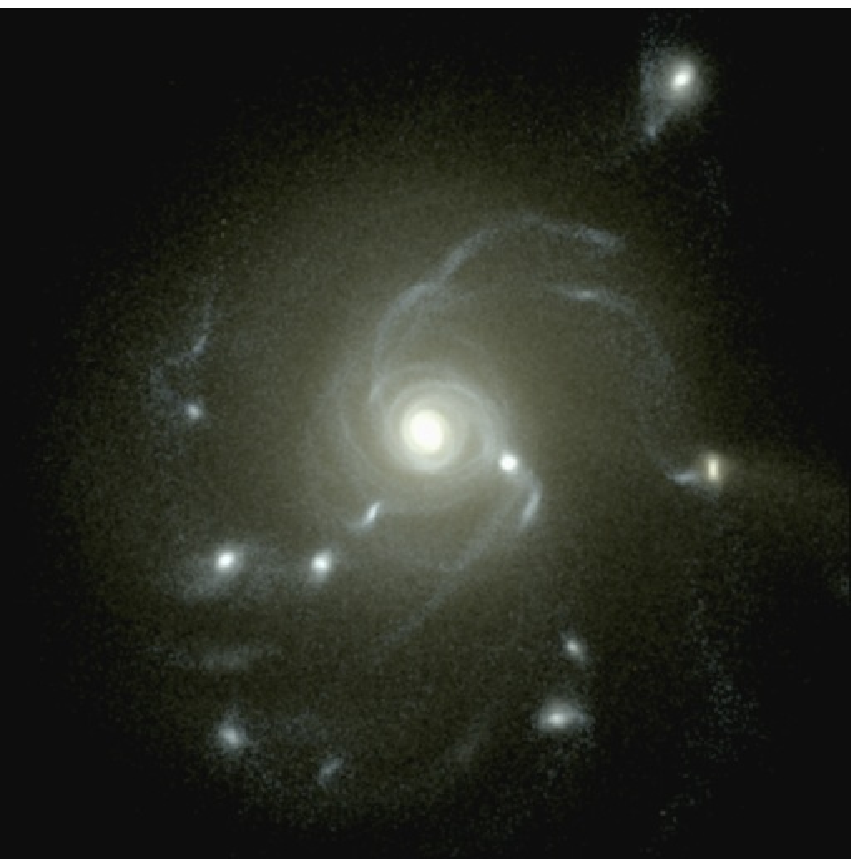}}}
  \centering{\resizebox*{!}{5.2cm}{\includegraphics{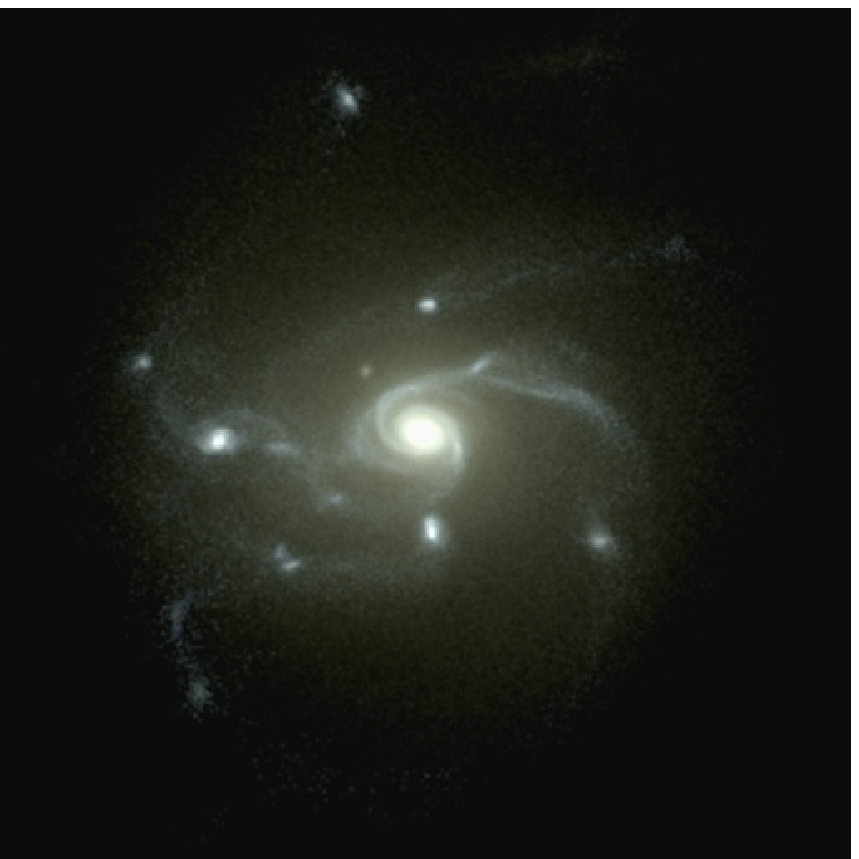}}}
  \centering{\resizebox*{!}{5.2cm}{\includegraphics{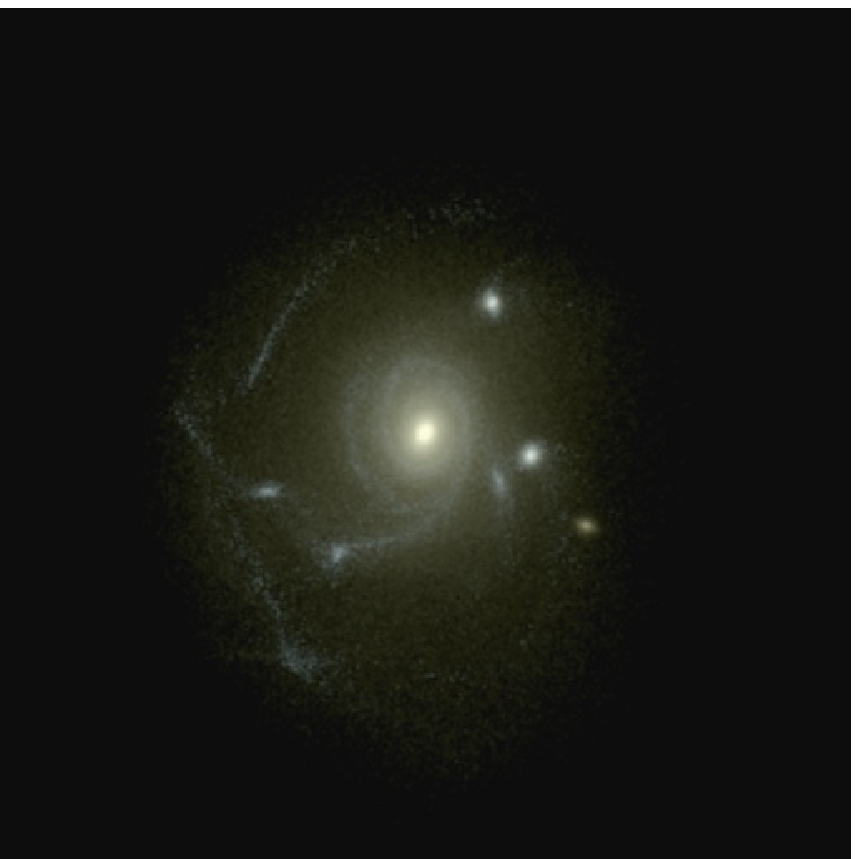}}}\hspace{0.8cm}
  \caption{Images of the gas density (top), and the stellar luminosity in the SDSS ugr filter bands (bottom) of the central galaxy seen face on at $z=6$ for the \nofeed~case (left), the \noAGN~case (middle), and the \AGN~case (right). The image sizes are 5 kpc length and deep. The \nofeed/\noAGN~cases have comparable gas densities, stellar intensities and number of clumps, while the \AGN~case is different. }
    \label{fig:starsmap}
\end{figure*}

We use the same ``canonical'' AGN feedback modeling employed in~\cite{duboisetal12agnmodel}.
BHs are created at loci where the gas density is larger than the density threshold for star formation $\rho_0$ with an initial seed mass of $10^5\, \rm M_\odot$.
We assume that very massive BHs form from the direct collapse of the low-angular momentum gas in primordial halos~\citep{begelmanetal06}.
In order to avoid the formation of multiple BHs in the same galaxy, BHs are not allowed to form at distances smaller than 5 kpc from any other BH particle.
The accretion rate onto BHs follows the Bondi-Hoyle-Lyttleton~\citep{bondi52} rate
$\dot M_{\rm BH}=4\pi \alpha G^2 M_{\rm BH}^2 \bar \rho / (\bar c_s^2+\bar u^2) ^{3/2},$
where $M_{\rm BH}$ is the BH mass, $\bar \rho$ is the average gas density, $\bar c_s$ is the average sound speed, $\bar u$ is the average gas velocity relative to the BH velocity, and $\alpha$ is a dimensionless boost factor with $\alpha=(\rho/\rho_0)^2$ when $\rho>\rho_0$ and $\alpha=1$ otherwise~\citep{booth&schaye09} in order to account for our inability to capture the colder and higher density regions of the ISM.
The effective accretion rate onto BHs is capped at the Eddington accretion rate:
$\dot M_{\rm Edd}=4\pi G M_{\rm BH}m_{\rm p} / (\epsilon_{\rm r} \sigma_{\rm T} c),$
where $\sigma_{\rm T}$ is the Thompson cross-section, $c$ is the speed of light, $m_{\rm p}$ is the proton mass, and $\epsilon_{\rm r}$ is the radiative efficiency, assumed to be equal to $\epsilon_{\rm r}=0.1$ for the \cite{shakura&sunyaev73} accretion onto a Schwarzschild BH.
In order to avoid spurious oscillations of the BH in the gravitational potential well due to external perturbations and finite resolution effects, we introduce a drag force that mimics the dynamical friction exerted by the gas onto a massive particle.
This dynamical friction is proportional to $F_{\rm DF}=f_{\rm gas} 4 \pi \alpha \rho (G M_{\rm BH}/\bar c_s)^2$, where $f_{\rm gas}$ is a fudge factor whose value is between 0 and 2 and is a function of the mach number ${\mathcal M}=\bar u/\bar c_s<1$~\citep{ostriker99, chaponetal11}, and where we introduce the boost factor $\alpha$ for the same reasons than stated above.

The AGN feedback is a combination of two different modes, the so-called \emph{radio} mode operating when $\chi=\dot M_{\rm BH}/\dot M_{\rm Edd}< 0.01$ and the \emph{quasar} mode active otherwise.
The quasar mode corresponds to an isotropic injection of thermal energy into the gas within a sphere of radius $\Delta x$, at an energy deposition rate: $\dot E_{\rm AGN}=\epsilon_{\rm f} \epsilon_{\rm r} \dot M_{\rm BH}c^2$,
where $\epsilon_{\rm f}=0.15$ for the quasar mode is a free parameter chosen to reproduce the $M_{\rm BH}$-$M_{\rm b}$, $M_{\rm BH}$-$\sigma_{\rm b}$, and BH density in our local Universe (see \citealp{duboisetal12agnmodel}).
At low accretion rates on the other hand, the radio mode deposits the AGN feedback energy into a bipolar outflow with a jet velocity of $10^4\,\rm km/s$ into a cylinder with a cross-section of radius $\Delta x$ and height $2 \, \Delta x$ following~\cite{ommaetal04} (more details about the jet implementation are given in~\citealp{duboisetal10}).
The efficiency of the radio mode is larger with $\epsilon_{\rm f}=1$.
We insist that we are taking the parameters of the AGN model at face-value (as we do for the parameters of star formation and SN feedback) calibrated on lower resolution hydro simulations that reproduce the~\cite{magorrianetal98} relations at $z=0$.

We are using three sets of simulations to asses the role of the different modes of feedback, focusing on the role of AGN feedback on galactic evolution and the mass content in the halo.
\begin{itemize}
\item{\nofeed: the reference simulation does not include any type of feedback. Only gas cooling, background UV heating and star formation are allowed.}
\item{\noAGN: the second simulation includes all the physics of the \nofeed~run in addition to feedback from type II SNe, but no AGN feedback.}
\item{\AGN: the final simulation includes all the physics of the \noAGN~run and feedback from AGN in a radio/quasar mode as described above.}
\end{itemize}

We also complemented these runs with two other simulations which include AGN feedback with different flavors: one where only the \emph{radio} (jet) mode is active to test the effect of the anisotropy of the energy injection, and another with the dual radio/quasar mode but with the AGN efficiency $\epsilon_{\rm f}$ reduced by a factor of 15 to test how this affects the energy released by BHs.
We do not detail the results from these simulation runs, but will punctually use them to reinforce the robustness of our findings concerning the impact of AGN feedback on structure formation (see Appendices~\ref{appendix:jet} and~\ref{appendix:efficiency}).

We performed some additional simulations to test the impact of changing the resolution in DM mass and in minimum cell size on our final results.
We find no significant difference with the results obtained using our canonical resolution, even though the resolution elements are changed by one order of magnitude (in DM mass and minimum cell size, see Appendix~\ref{appendix:resolution} for further details). 

\begin{figure}
  \centering{\resizebox*{!}{6.cm}{\includegraphics{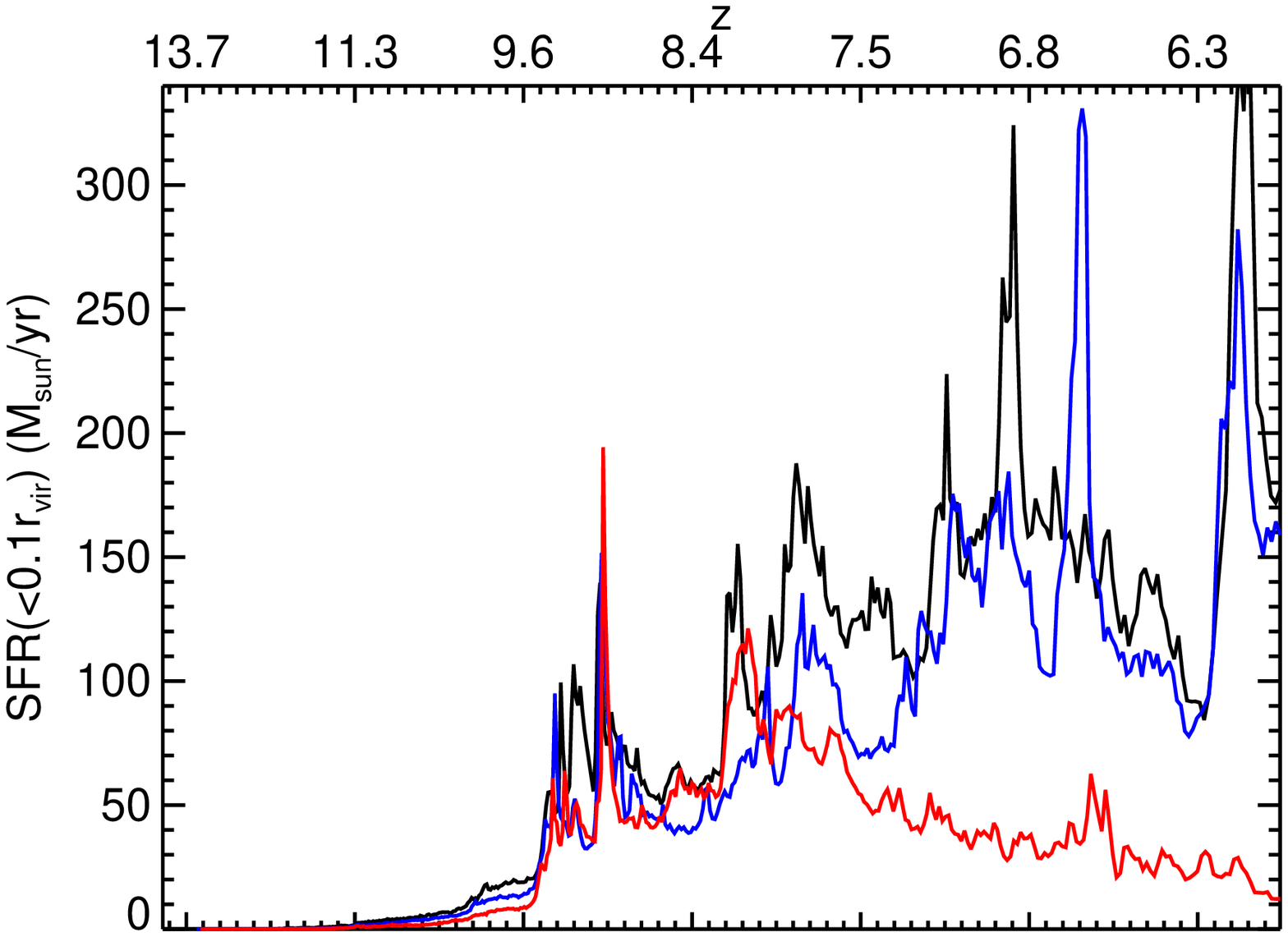}}}\vspace{-1.3cm}
  \centering{\resizebox*{!}{6.cm}{\includegraphics{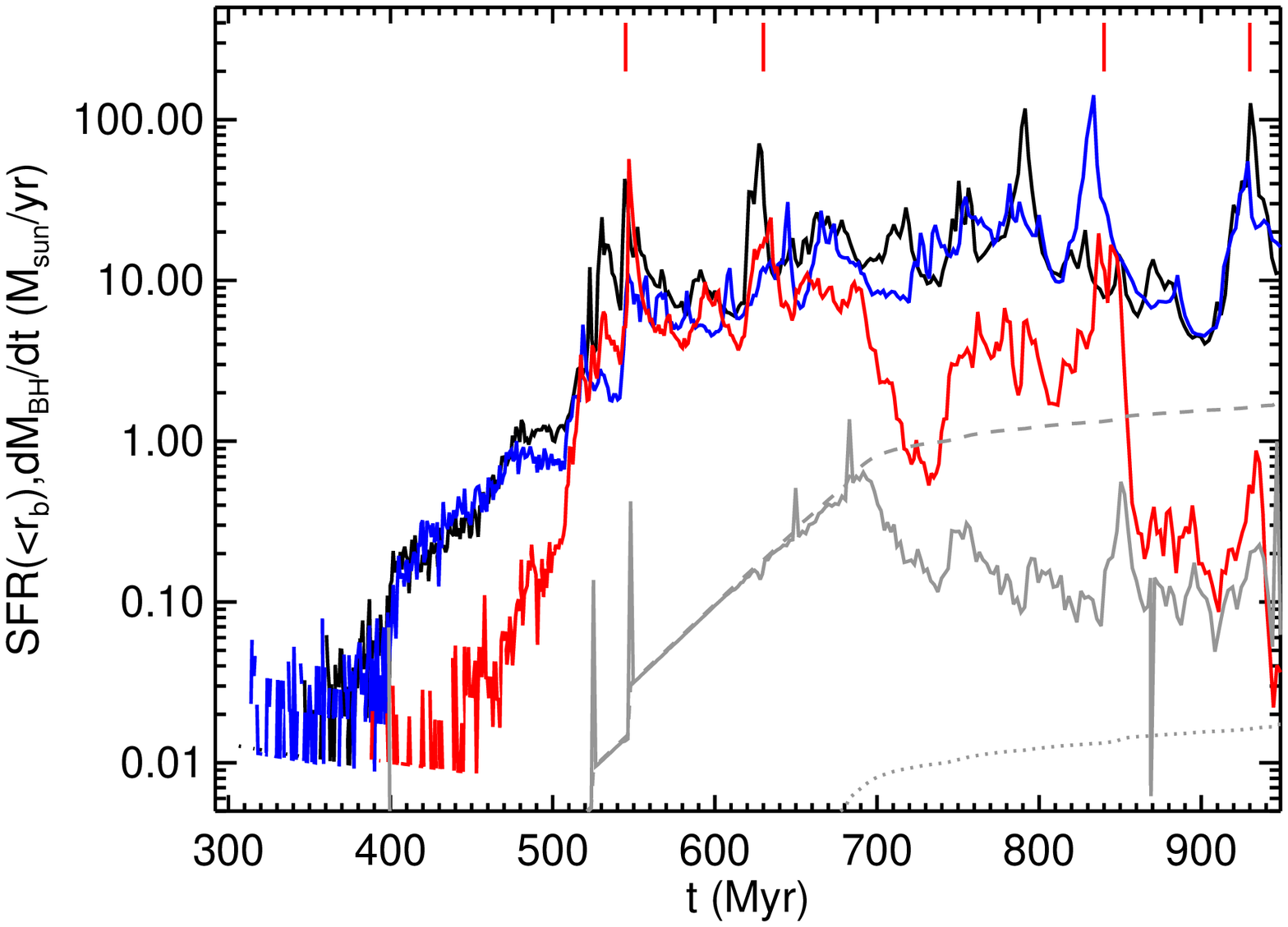}}}
  \caption{\emph{Top}: SFR history using the stars contained within $0.1\, r_{\rm vir}$ for the \nofeed~case (black), the \noAGN~case (blue), and for the \AGN~case (red). \emph{Bottom}: SFR within $r_{\rm b}=50 \, \rm pc$ (same color coding as top panel), and BH accretion rate (grey solid), Eddington accretion rate (grey dashed), and 1 per cent of Eddington accretion rate (grey dotted). Merger events for the \AGN~run are marked with vertical red lines. SFRs of the \nofeed/\noAGN~cases show almost no difference, while the SFR in the \AGN~case is significantly reduced when the central BH exhibits a departure from its Eddington-limited growth after $z=8$.}
    \label{fig:sfr}
\end{figure}

\begin{figure}
  \centering{\resizebox*{!}{6.0cm}{\includegraphics{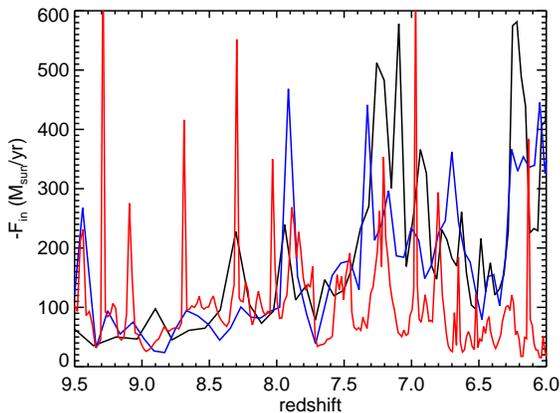}}}
  \caption{Mass accretion rate at $0.1 \, r_{\rm vir}$ as a function of redshift for the \nofeed~case (black), \noAGN~case (blue), \AGN~case (red). Smooth accretion onto the galaxy is reduced in the \AGN~case after $z=8$, but mergers continue to proceed.}
    \label{fig:fluxoutgal}
\end{figure}

\section{Quenching  star formation and  catastrophic gas cooling}
\label{section:bulge}

Fig.~\ref{fig:starsmap} shows the gas density and the stellar luminosity (SDSS ugr filters) images of the central galaxy at $z=6$ for the three different runs, revealing in each case, a massive bright bulge component with multiple clumps in rotation with the disc component.
However the \nofeed~and the \noAGN~simulations do not show large differences: they share a similar number of gas/stellar clumps with similar densities/intensities.
In contrast, as a consequence of the impact of AGN feedback on the baryon content of the galaxy, the \AGN~case has fewer clumps and less dense gas available in the galaxy.
We also note that recent AGN activity in the bulge creates a hole in the gas distribution in the centre of the galaxy in the \AGN~run.

The measured star formation rates (SFRs) in the galaxy for the stars enclosed within $0.1 \, r_{\rm vir}$ at $z=6$, indicate that, without AGN, the galaxy maintains high levels of star formation ($>100 \,\rm M_\odot\, yr^{-1}$) (see Fig.~\ref{fig:sfr}).
Feedback from SNe has only a marginal effect on the amount of stars formed in the galaxy.
This is also the case if the SFR measurement is done for the stars in the bulge of the galaxy ($r < r_{\rm b}$ where $r_{\rm b}=50 \, \rm pc$).
The bulge and disc components are decomposed by fitting two exponentially decreasing profiles $\Sigma_i (r)=\Sigma_{i,0}\exp(-r/r_i)$\footnote{Bulges are more usually defined by a S\'ersic profile with an index depending on its morphological classification (bulge or pseudo-bulge). We do the simple assumption that our simulated stellar bulge profiles can be described by a S\'ersic profile and index of 1, corresponding to an exponentially decreasing profile.} plus one gaussian profile (for the bar component if any) on the stellar surface densities.
Masses of the bulge and of the disc are defined as $M_{\rm i}=\Sigma_{i,0} 2\pi r_i^2$.
The measured bulge stellar mass is the same for the \nofeed~run and the \noAGN~run ($M_{\rm b,nof,noA}=1.9\times10^{10}\,\rm M_\odot$ at $z=6$), while the disc stellar mass is reduced by 20 per cent due to the presence of SNe (from $M_{\rm d,nof}=3.0 \times10^{10}\,\rm M_\odot$ to $M_{\rm d,noA}=2.4\times 10^{10}\,\rm M_\odot$).

As already suggested qualitatively by Fig.~\ref{fig:starsmap}, the SFR measurement for the \AGN~run confirms that AGN feedback has a dramatic impact on the amount of stars formed in the galaxy.
The global SFR is reduced to a few $10 \,\rm M_\odot\,yr^{-1}$, and the SFR in the bulge is almost completely quenched to less than $1 \,\rm M_\odot\,yr^{-1}$, two orders of magnitude below the \noAGN~case.
Indeed, it has severe consequences for the bulge stellar mass that is reduced by a factor 3 with $M_{\rm b,A}=6.2 \times 10^{9}\,\rm M_\odot$ at $z=6$.
The bulge, despite the impact of AGN feedback, is more compact than the most compact bulges of comparable stellar mass at $1<z<3$~\citep{bruceetal12}, indicating that bulges follow a strong evolution with redshift and are more compact in the distant Universe~\citep{daddietal05, khochfar&silk06, cimattietal08}.
The disc stellar mass is also reduced to $M_{\rm d,A}=1.2 \times 10^{10}\,\rm M_\odot$, mainly because the AGN clears the centre of the halo, causing the accretion of new material to proceed more quietly, while for the \noAGN~run, energy from SNe is not sufficient  to significantly impact  the accretion rate at $0.1\, r_{\rm vir}$ (see Fig.~\ref{fig:fluxoutgal}).
Indeed after $z=8$ the mass inflow onto the galaxy proceeds with a smooth component, corresponding to the accretion of cold streams,  at a few $10\, \rm M_\odot\,yr^{-1}$ for the \AGN~run, as opposed to $>100\, \rm M_\odot\,yr^{-1}$ for the \nofeed/\noAGN simulation.
Bursty episodes mark the accretion of small satellites.
In summary, AGN are able to substantially reduce the smooth part of the accretion flow, but the bursty episodes of accretion persist.

\begin{figure*}
  \centering{\resizebox*{!}{5cm}{\includegraphics{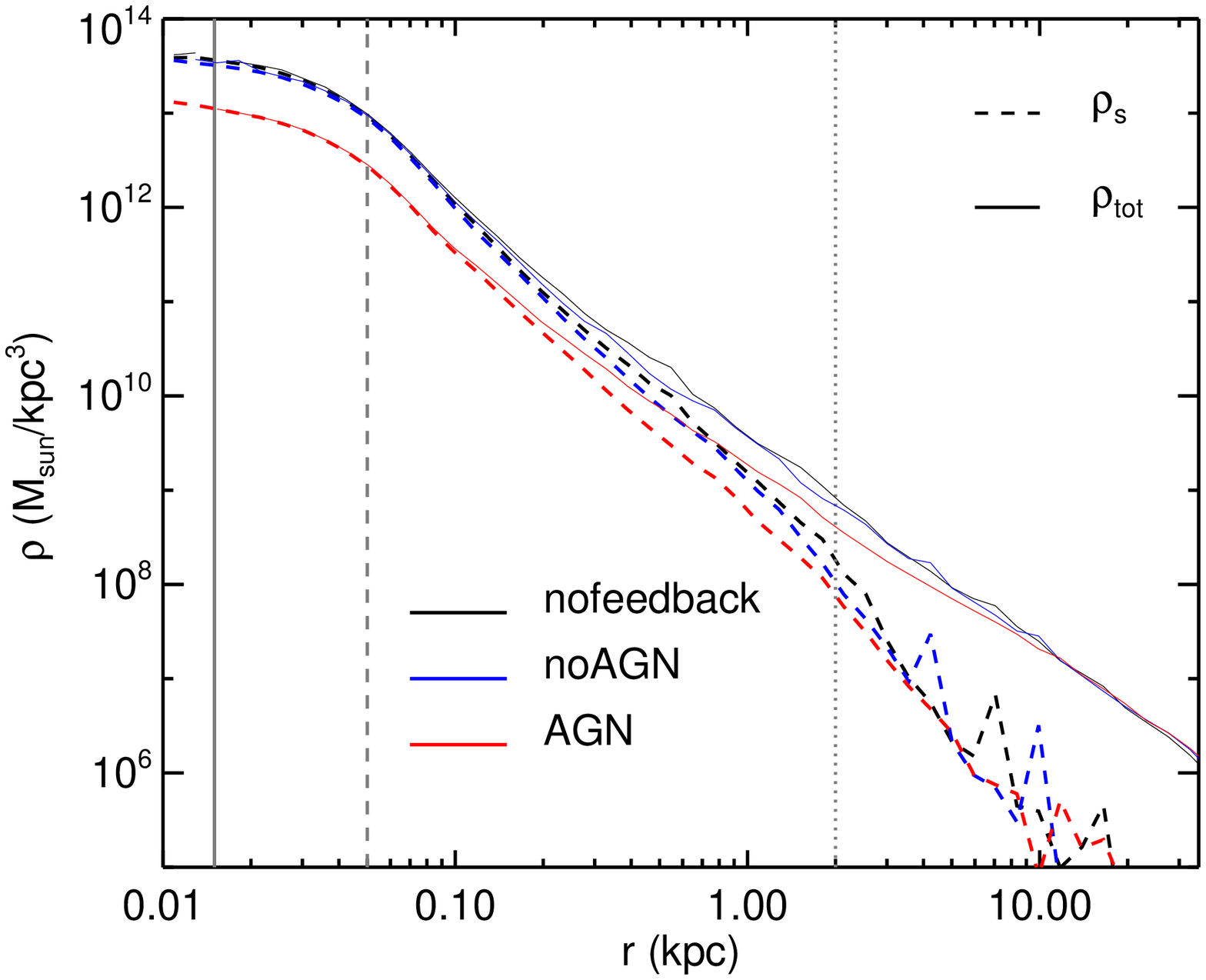}}}\hspace{-1cm}
  \centering{\resizebox*{!}{5cm}{\includegraphics{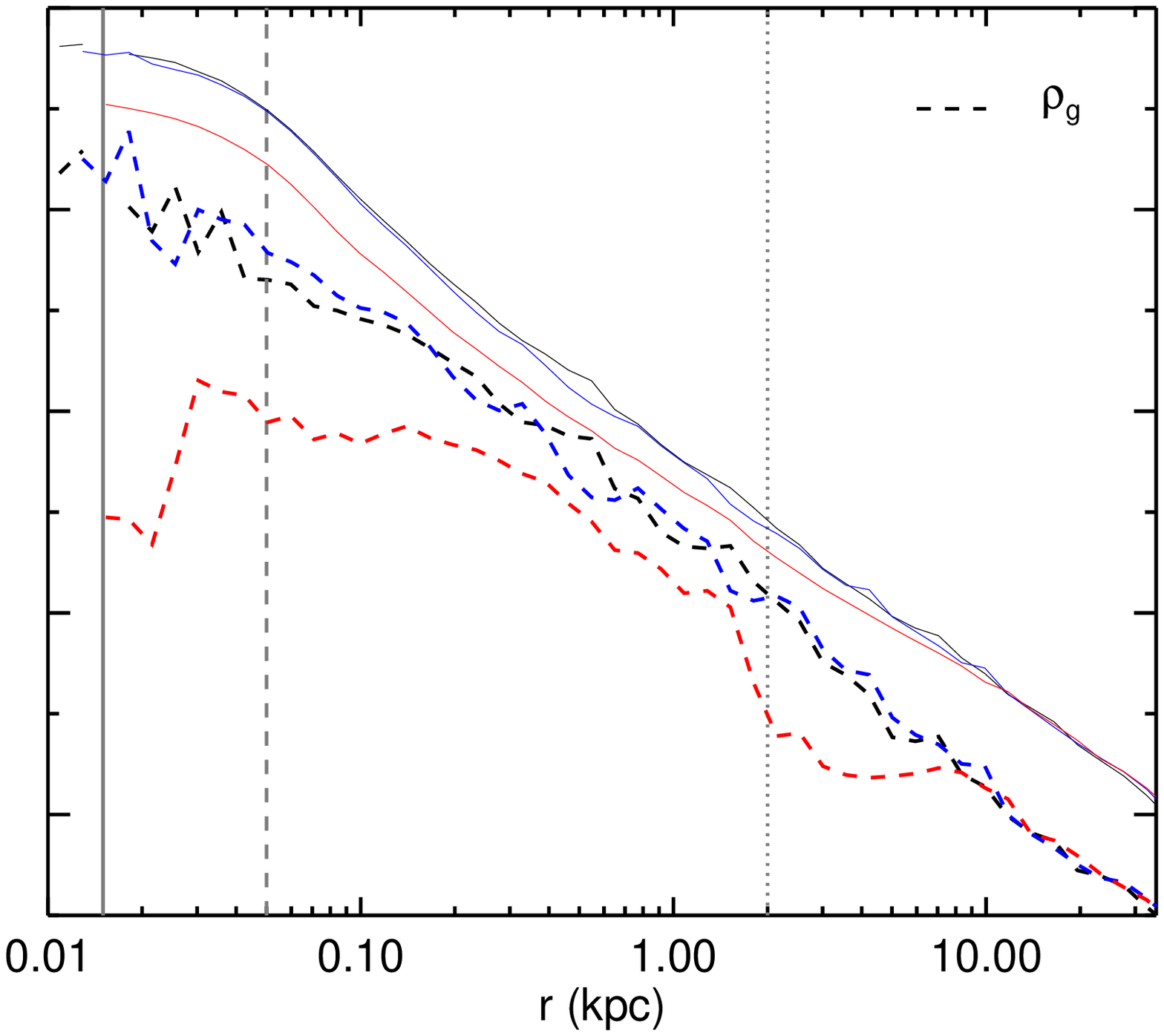}}}\hspace{-1cm}
  \centering{\resizebox*{!}{5cm}{\includegraphics{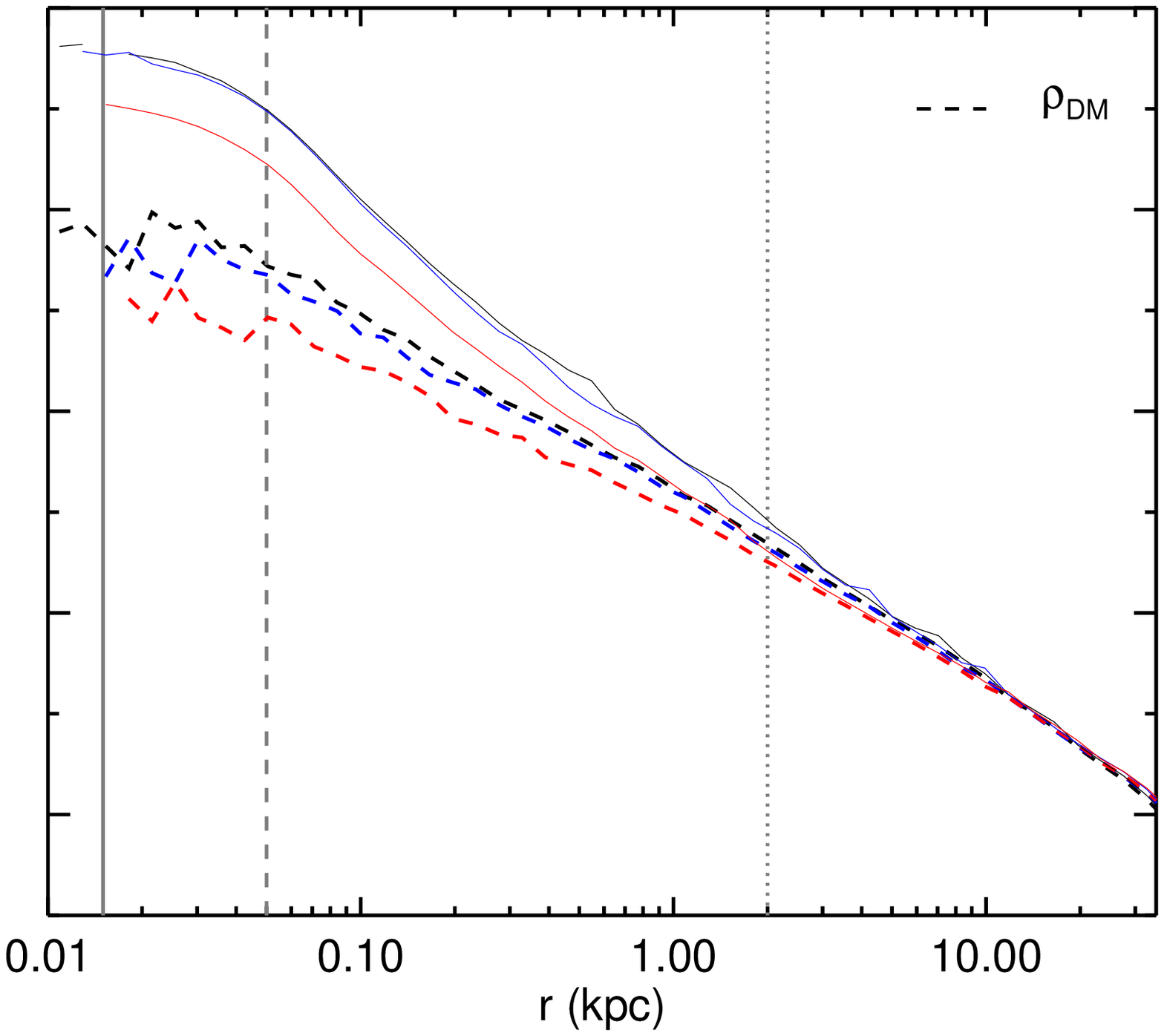}}}
  \caption{Density profiles of all the components (gas$+$stars$+$DM) (solid lines), stars (left panel dashed), gas (middle panel dashed), and DM (right panel dashed) for the \nofeed~case (black), the \noAGN~case (blue), and the \AGN~case (red) at $z=6$.  The vertical grey solid line indicates the resolution limit, the vertical grey dashed lines indicates the scale radius of the bulge, and the vertical grey dotted line the scale radius of the disc.}
    \label{fig:rhovsr}
\end{figure*}

The consequences of AGN feedback on the mass profiles of baryons in the galaxy are also dramatic (Fig.~\ref{fig:rhovsr}).
The stellar and gas densities are reduced both in the bulge and the disc with the formation of a gap in the gas distribution within the bulge due to the outflow produced by the central BH.
Again, the strongest impact of the AGN is on the bulge component, but the disc is also significantly altered because the accretion onto the galaxy is quenched efficiently.
A natural outcome is that the peak of the circular velocity is reduced from 900 to 500 $\rm km\, s^{-1}$, together with a reduced velocity dispersion of the stars within the bulge from 450 to 250 $\rm km \, s^{-1}$, consistent with velocity dispersions observed in $z=6$ galaxies inferred from velocity widths of CO lines~\citep{wangetal10}.
Note also the deficit of gas between the disc radius at $r=2$~kpc and 10~kpc that is due to efficient ejection of gas around the galaxy.
The fact that the stellar and gas distribution in the \noAGN~case are similar to their \nofeed~counterparts, even in the disc component, means that SNe do not efficiently remove gas or quench the gas infall onto the galaxy, and that with similar gas content in the galaxy, identical stellar distributions are found.
The ability of AGN feedback to prevent the baryons from collapsing and/or to remove the gas from the galaxy early-on weakens the adiabatic contraction of the dark matter particles~\citep{blumenthaletal86} and is therefore key to understanding mass density distributions in galaxy clusters~\citep{duboisetal10, duboisetal11, teyssieretal11, martizzietal12b}.
Indeed, we see that with AGN feedback, the density profiles of gas, stars and DM are shallower inside the galaxy, as AGN feedback  prevents the cold gas from collapsing too much.
The measured mass of DM in the centre of the halo within $0.1\, r_{\rm vir}$ has been reduced by 35 per cent to $M_{\rm DM, A}(0.1\, r_{\rm vir})=3.8\times 10^{10} \, \rm M_\odot$ in the \AGN~case compared to the mass $M_{\rm DM, nof}(0.1\, r_{\rm vir})=6.0\times 10^{10} \, \rm M_\odot$ in the \nofeed~case at $z=6$ ($M_{\rm DM, noA}(0.1\, r_{\rm vir})=5.4\times 10^{10} \, \rm M_\odot$ in the \noAGN~case). 

\begin{figure*}
  \centering{\resizebox*{!}{7.3cm}{\includegraphics{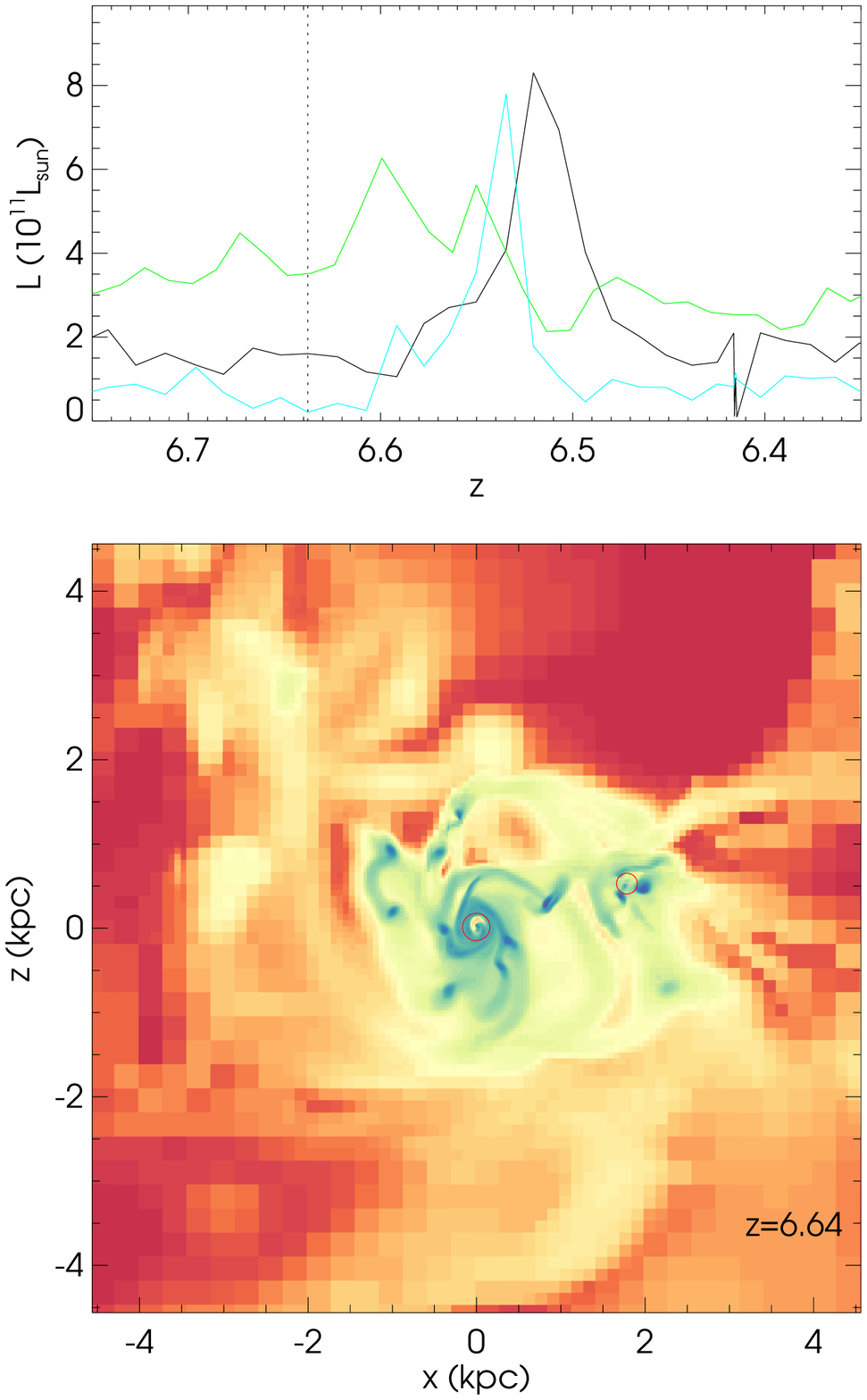}}}\hspace{-0.5cm}
  \centering{\resizebox*{!}{7.3cm}{\includegraphics{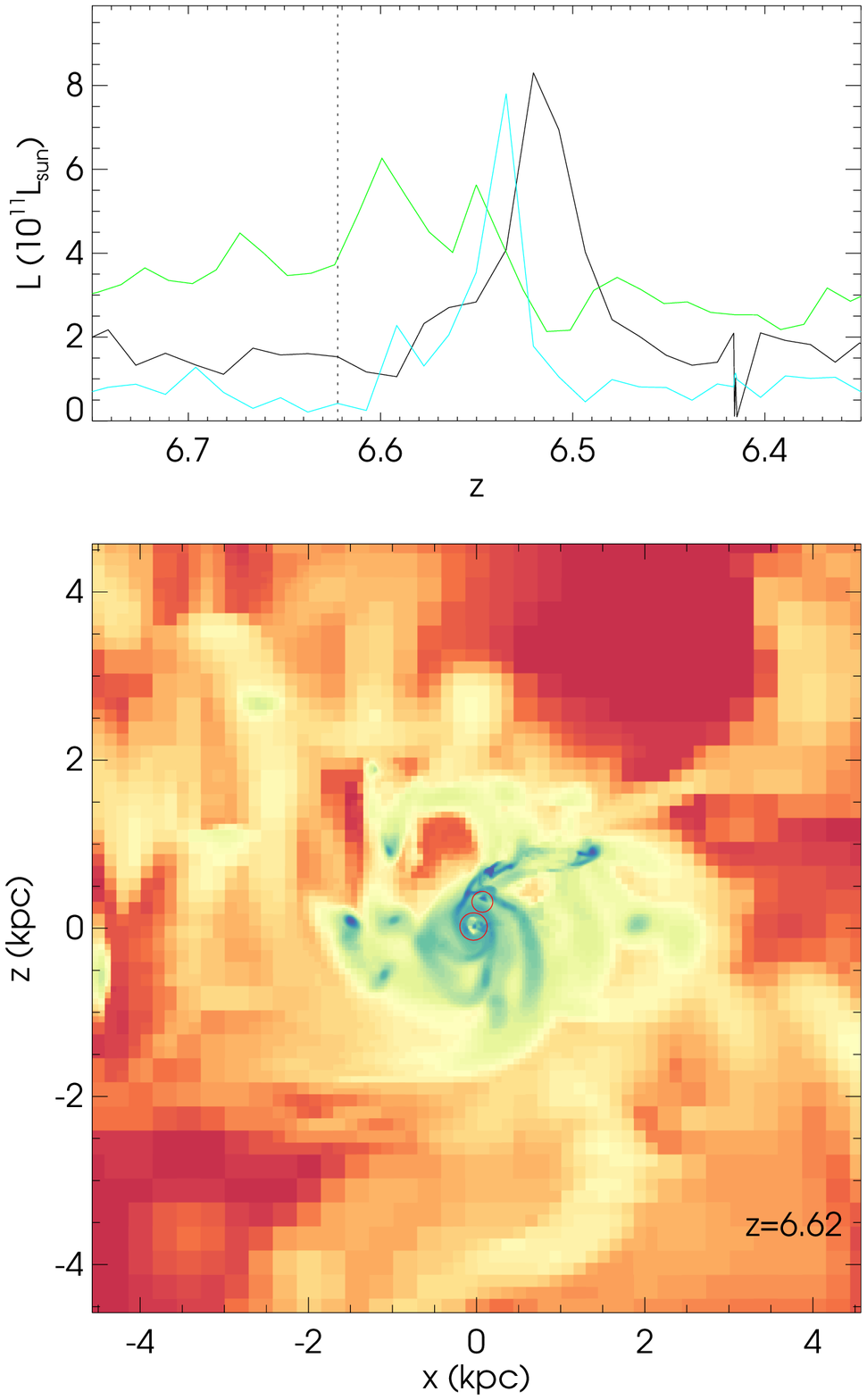}}}\hspace{-0.5cm}
  \centering{\resizebox*{!}{7.3cm}{\includegraphics{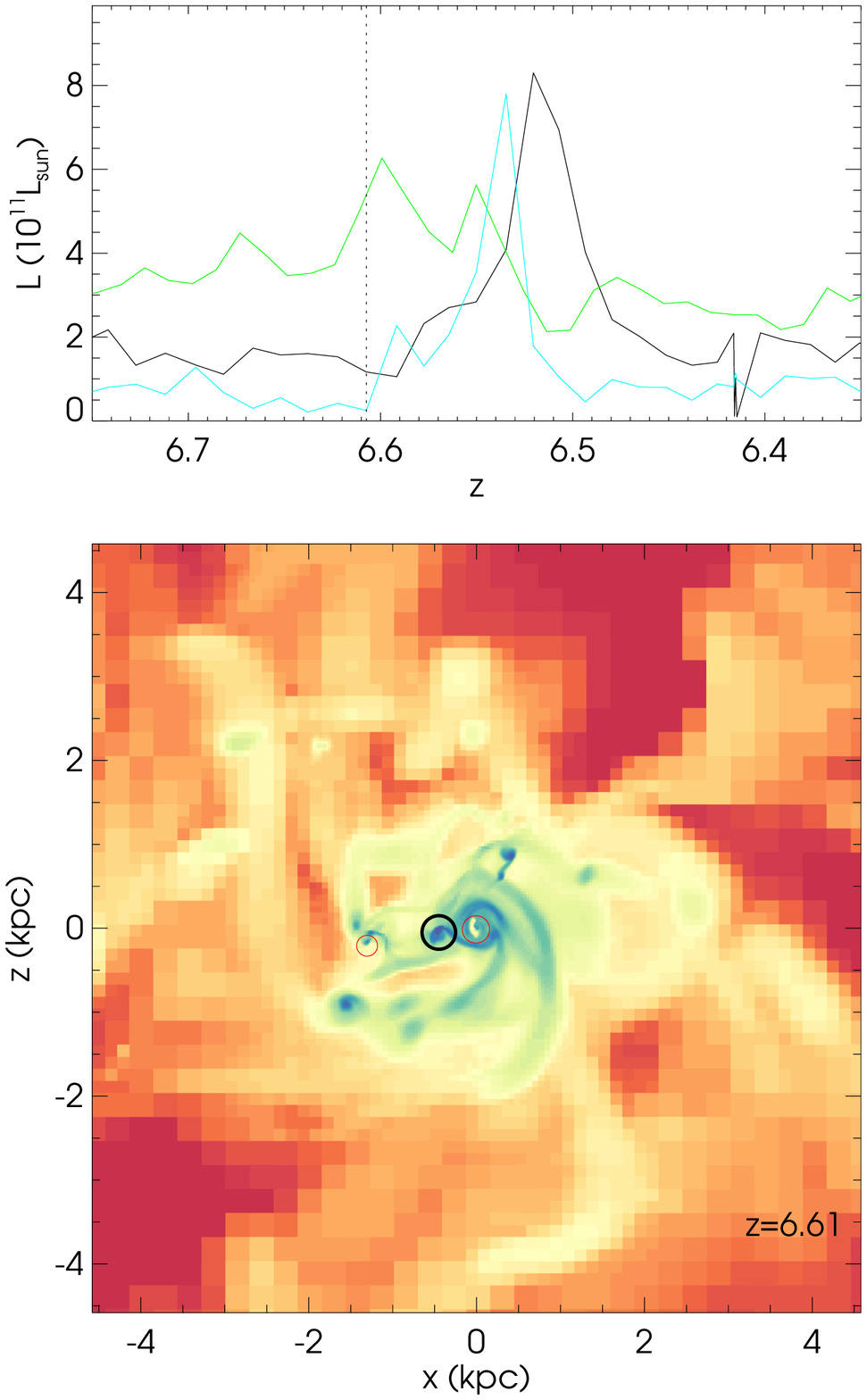}}}\hspace{-0.5cm}
  \centering{\resizebox*{!}{7.3cm}{\includegraphics{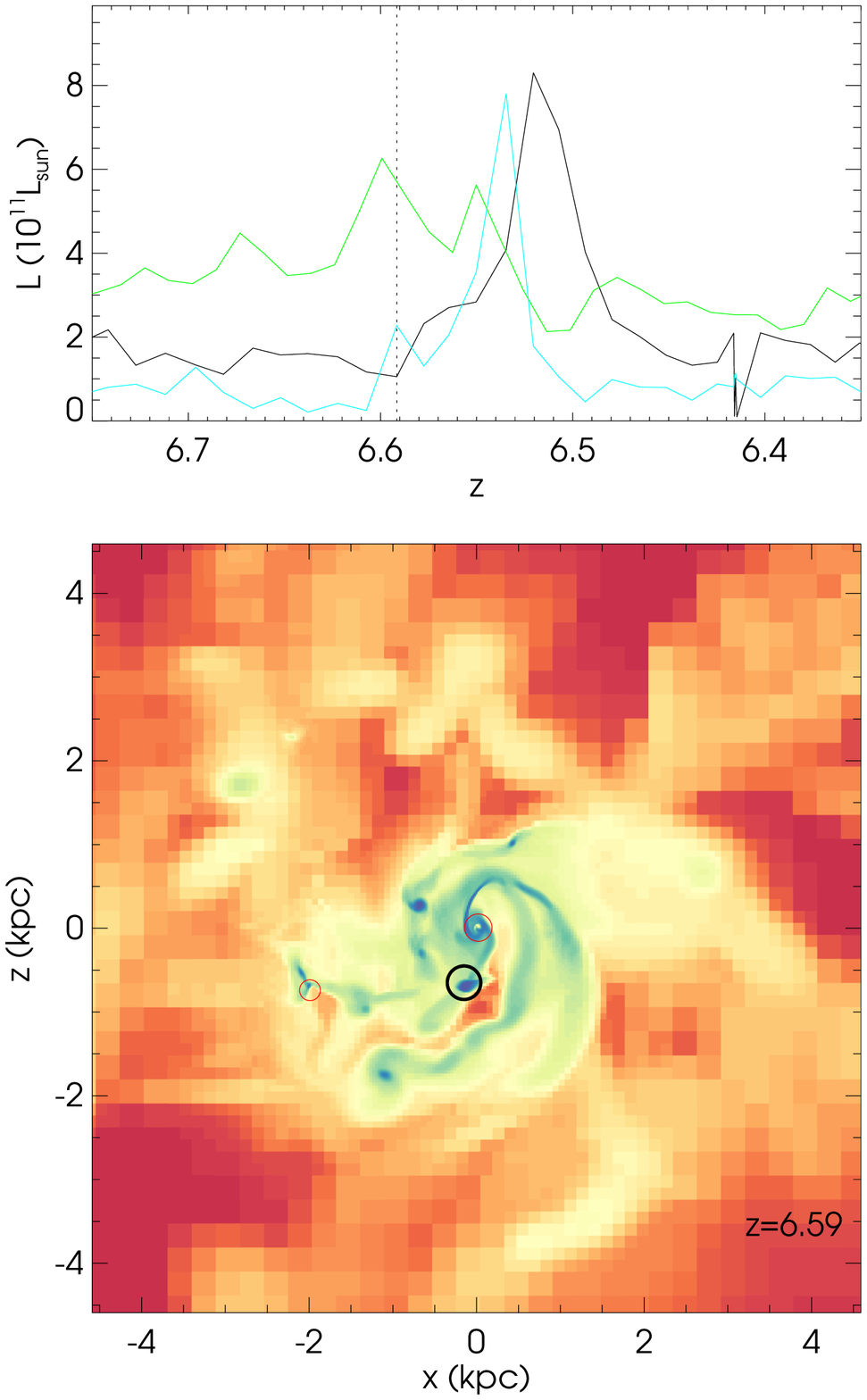}}}
  \centering{\resizebox*{!}{4.8cm}{\includegraphics{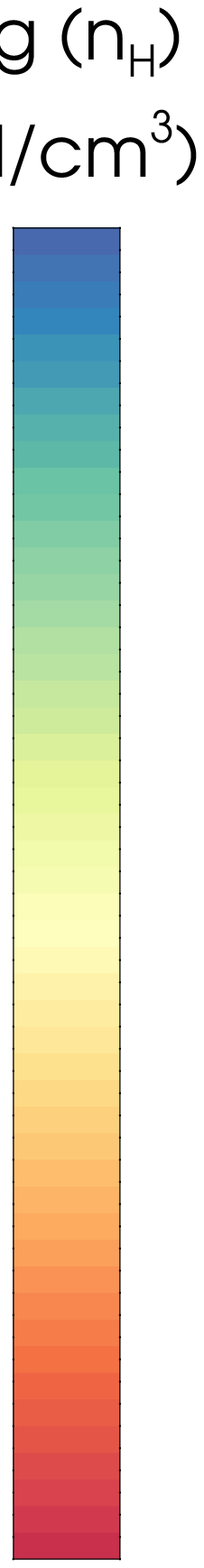}}}\vspace{-0.5cm}
  \centering{\resizebox*{!}{7.3cm}{\includegraphics{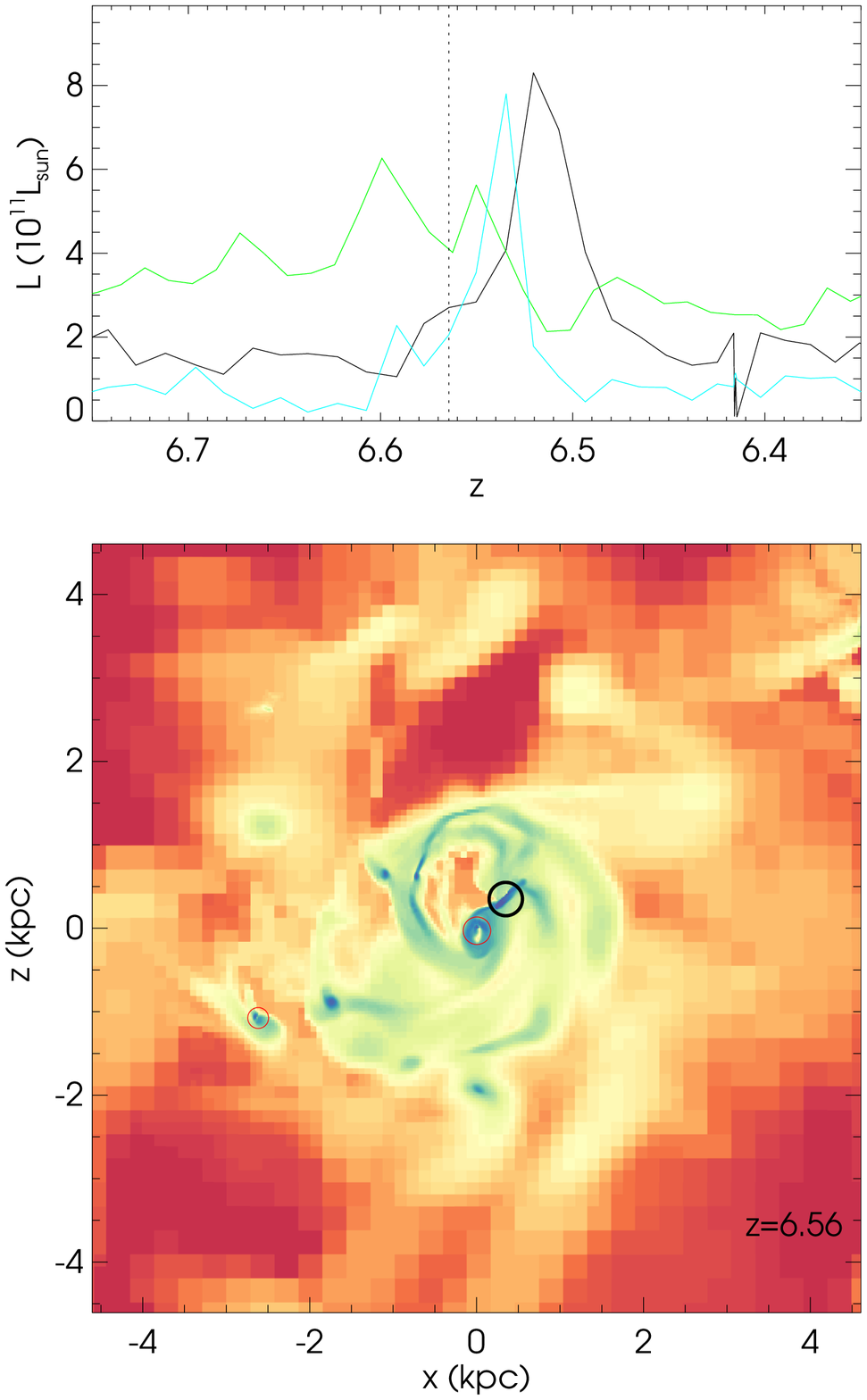}}}\hspace{-0.5cm}
  \centering{\resizebox*{!}{7.3cm}{\includegraphics{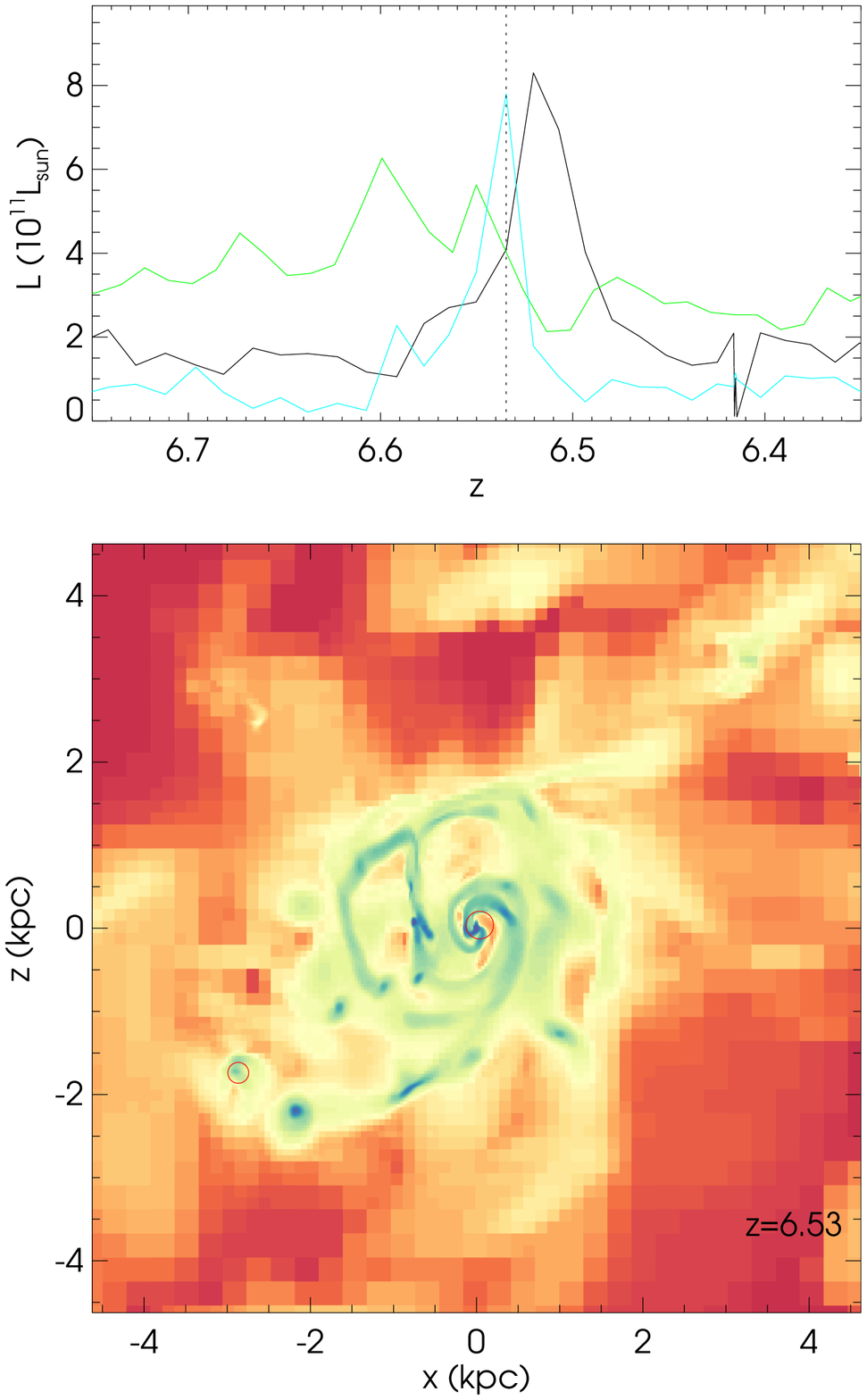}}}\hspace{-0.5cm}
  \centering{\resizebox*{!}{7.3cm}{\includegraphics{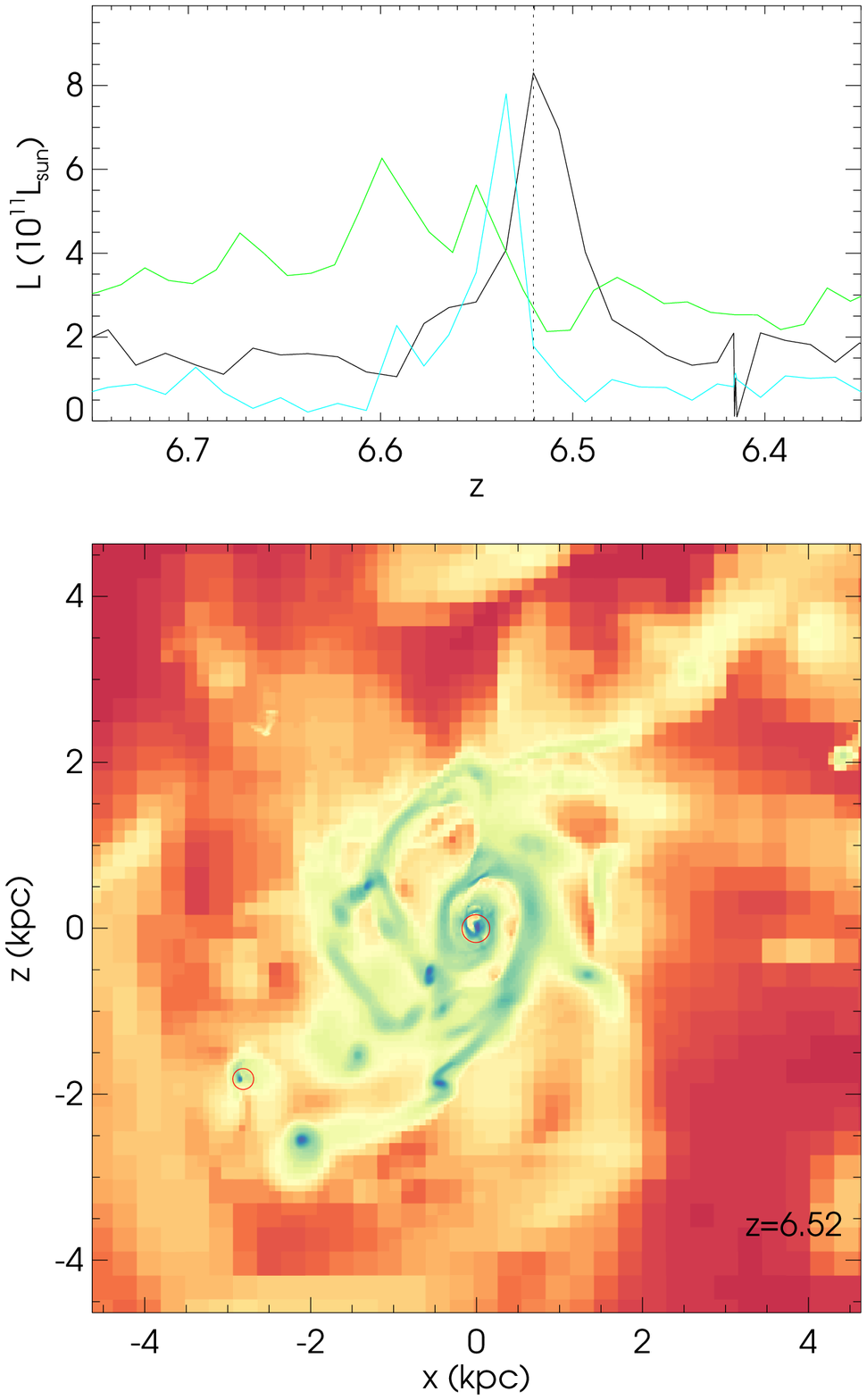}}}\hspace{-0.5cm}
  \centering{\resizebox*{!}{7.3cm}{\includegraphics{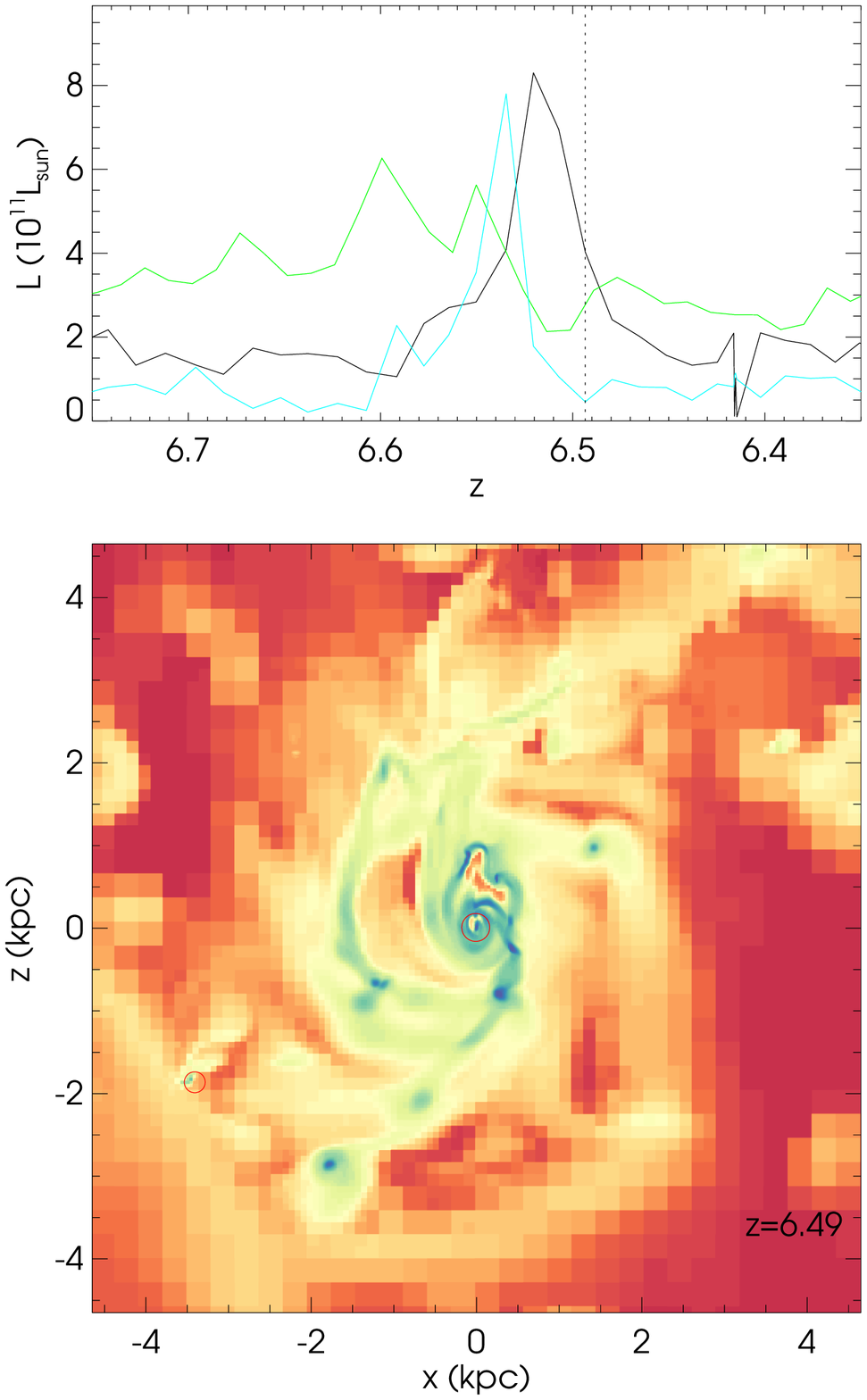}}}
  \centering{\resizebox*{!}{4.8cm}{\includegraphics{./fig/colortab_fig5.ps}}}\\
  \caption{Time sequence (as labelled) around a peak of starburst and quasar activity triggered by a merger of 1:4 ratio at $z\simeq6.5$ of the gas density of the central galaxy. Red circles correspond to the position of the two most massive BHs. The projections of the images are 9 kpc deep. The panels on top of each image show the bolometric luminosity of the central BH in $\rm L_\odot$ (black) with the SFR (green) of the central galaxy and the scaled mass accretion rate onto the bulge measured at 80 pc (blue). The SFR and mass accretion rate values in $\rm M_\odot\, yr^{-1}$ are divided by a factor 10 and 30 respectively for a better comparison to the luminosity of the central BH. The burst of SFR happens at the passage of the satellite at the apocentre. The rapid quasar activity is triggered by the enhanced migration of a massive clump of dense gas (black circles) whose motion is perturbed by the presence of the satellite. Note that the peak of SFR and AGN activity are not synchronized. The final merger between the two galaxies is complete at $z=6.2$ .}
    \label{fig:timeseq_merger}
\end{figure*}

\section{What drives the AGN activity?}
\label{section:energy}

According to Fig.~\ref{fig:sfr}, up to 600 Myrs after the Big Bang the SFRs for the \AGN~case and the \noAGN~case are comparable.
The onset of SFR quenching at $z\simeq 7.5$ coincides with the redshift when the BH stops accreting gas at the Eddington limit (see bottom panel of Fig.~\ref{fig:sfr}).
After $z=8$, the growth of the central BH is regulated by its own \emph{quasar} mode feedback which can efficiently remove gas from the galaxy.
The strong outflows produced by the central quasar prevent gas accretion onto the galaxy (see Fig.~\ref{fig:fluxoutgal}), thereby starving the gas reservoir available for star formation.
The accretion of gas for the \AGN~case looks spikier compared to the \noAGN~case or \nofeed~case due to efficient removal of the diffuse component surrounding the galaxy with low ram-pressure, while the dense star forming gas continues to accrete onto the galaxy (see also Fig.~\ref{fig:fxnrvir}).
Because BH accretion rates hover at around a tenth of the Eddington rate, the AGN feedback proceeds in \emph{quasar} mode all the way down to $z=6$. Recall that in this mode energy is deposited isotropically.
However due to the presence of a dense disc component surrounding the bulge of the galaxy, the AGN launches bipolar outflows with large opening angles from the galaxy.

To test the importance of the directionality and collimation of the outflow on our results, we run a similar simulation to the \AGN~case where only the \emph{radio} (jet) mode is allowed.
Despite the absence of opening angle in the jet input, and its small size input ($\Delta x=15$~pc) compared to the extent of the galactic disc (2 kpc), we reach similar conclusions because the jet propagates as a light jet into the hot diffuse gas, and almost immediately spreads into a wide outflow (see Appendix~\ref{appendix:jet}).

Analysing the halo in the \nofeed~simulation,~\cite{duboisetal12angmom} have shown that the cold flows feed the central bulge directly until $z\simeq8$.
After this redshift, the filaments wrap around the galaxy because they gain angular momentum from the tidal field of the large scale cosmic structures~\citep{pichonetal11}.
This gives rise to the formation of a large disc component surrounding the bulge that is gravitationally unstable (with Toomre parameter of $Q\simeq 1$) causing disc instabilities and clump migration~\citep{elmegreenetal08, bournaudetal11} to bring large quantities of gas towards the bulge.
Interestingly, the central BH stops accreting at the Eddington rate at the moment when the gas accretion onto the bulge transitions from direct infall from cold streams to the feeding through disc instabilities (see Fig.~10 in~\citealp{duboisetal12angmom}). This suggests that the cold flows are directly responsible for maintaining sufficiently large amounts of gas in the bulge, so as to allow the BH to accrete at its maximum rate.

\begin{figure}
  \centering{\resizebox*{!}{4.cm}{\includegraphics{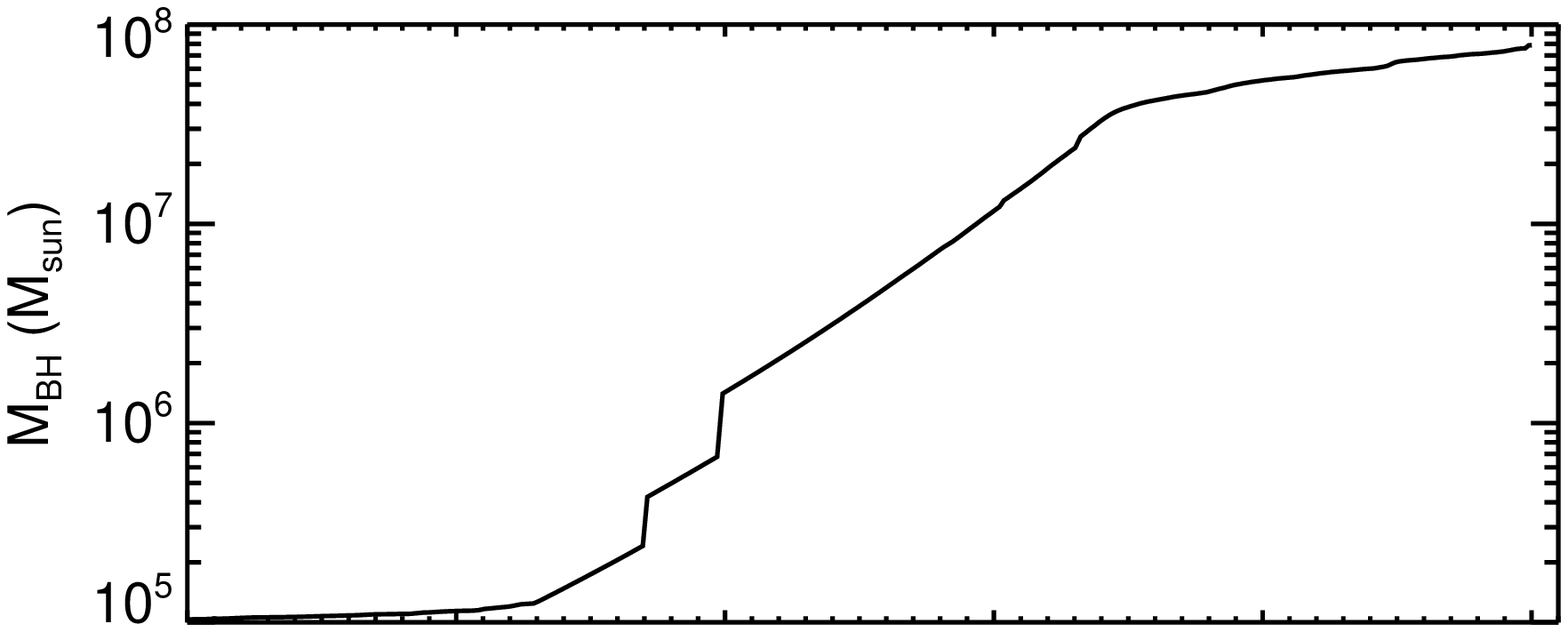}}}\vspace{-1.1cm}
  \centering{\resizebox*{!}{6.cm}{\includegraphics{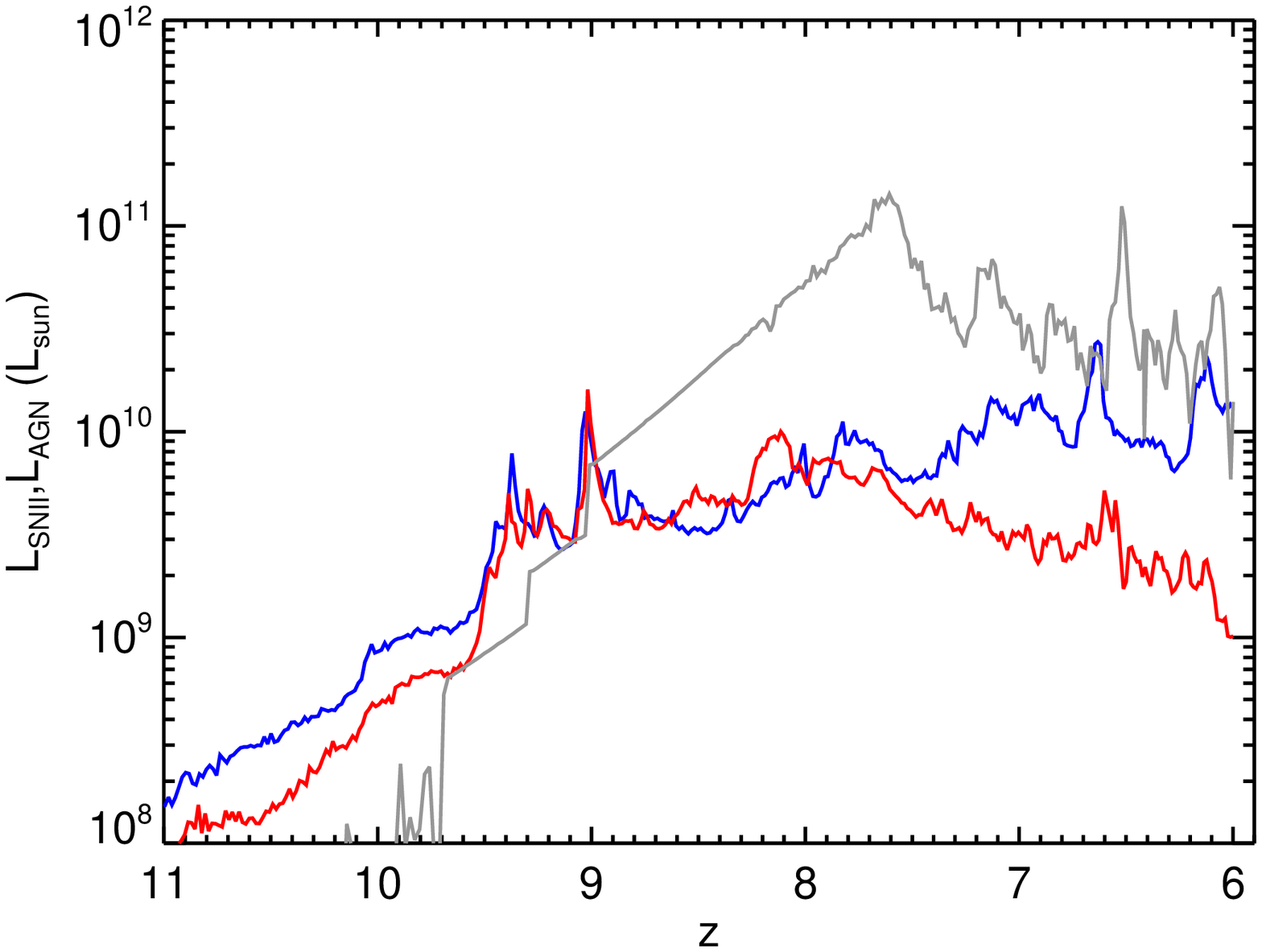}}}
  \caption{\emph{Top}: BH mass as a function of redshift. \emph{Bottom}: Feedback luminosity from type II SNe for the \noAGN~case (blue) and for the \AGN~case (red), and the feedback luminosity from the central AGN in the \AGN~run (grey). The energy output from the central AGN largely dominates the total energy budget from feedback.}
    \label{fig:lumvsz}
\end{figure}

Several mergers that correlate with short bursts of star formation can be identified in the SFR evolution (vertical red lines in Fig.~\ref{fig:sfr}).
The response of the SFR to such mergers is different depending on the presence or absence of AGN.
In the \nofeed~and \noAGN~cases, the SFR after a merging event reverts to the SFR levels present before the merger.
Thus, the outflow produced by SNe during the starburst is not strong enough to regulate the gas content in the galaxy.
In contrast, as soon as the BH enters a sub-Eddington mode of accretion (after $z\simeq8$) in the \AGN~case, mergers provoke a large drop of the SFR in the bulge, following the starburst: the SFR decreases by one order of magnitude below its pre-merger value.
This highlights the fact that mergers trigger large inflows of gas towards the centre of the merging galaxies and can efficiently feed the BHs with fresh gas that activates a strong period of AGN activity able to blow away the surrounding gas~\citep{dimatteoetal05, springeletal05, hopkinsetal06}.

The largest galaxy merger in the simulation occurring at $z\simeq 6.5$ with mass ratio 1:4 is illustrated in Fig.~\ref{fig:timeseq_merger} with a time sequence of the gas density along with the central quasar bolometric luminosity and SFR of the central galaxy.
We see that the peak of the quasar activity at $z=6.52$ is not synchronized with the peak of the starburst at $z=6.61$, corresponding to a time delay of 15 Myrs equal to the dynamical time of the galaxy $t_d=17$~Myr (computed for the enclosed mass within $0.1\, r_{\rm vir}$). 
The latter occurs when the satellite passes the apocentre, while the former is correlated with the sharp increase of the mass inflow rate to the bulge when a satellite exerts a global torque on the central galaxy.
An important point is that it is not because the two galaxies merge that the AGN activity peaks (the merger of the two galaxies is not complete by $z=6.2$), but rather because the satellite applies this strong torque on the disc of the central galaxy.
In response, the gas looses angular momentum and plunges radially towards the bulge, and clumps of gas undergo a faster migration towards the bulge.
This effect is illustrated with the clump circled in black  in Fig.~\ref{fig:timeseq_merger} with gas mass $M_{\rm g,C}=5\times 10^8\, \rm M_\odot$. 
The final capture of the clump corresponds to the burst in the bulge accretion rate followed by the peak of the AGN activity.
The idealized picture described in~\cite{bournaudetal11} where the accretion onto the central BH is triggered by the rapid migration of gas clumps into the bulge should therefore be supplemented by the following: external perturbations exert strong torques on the disc of the galaxy that further accelerate the accretion of clumps onto bulges.

Computing the energy budget involved in the different feedback processes helps us to explain why AGN feedback is able to unbind the gas from the galactic centre while SNe are extremely inefficient.
The total energy injected by the central BH through the quasar mode is $E_{\rm AGN}=\epsilon_f\epsilon_r M_{\rm BH} c^2\simeq 2 \times 10^{60}$~erg where we take $M_{\rm BH}$ to be the mass of the central BH  $M_{\rm BH}=8 \times 10^7 \, \rm M_\odot$ at $z=6$ (Fig.~\ref{fig:lumvsz}), whereas the energy liberated by SN explosions occurring in the bulge is $E_{\rm SN,A}=\eta_{\rm SN}M_{\rm b, A} (10^{51}\rm erg /10 M_\odot)\simeq 6 \times 10^{58}$~erg for the \AGN~case ($E_{\rm SN,noA}\simeq 2 \times 10^{59}$~erg for the \noAGN~case).
For comparison, the binding energy of the bulge is $E_{\rm b,A}\simeq G M_{\rm b}^2/r_{b}\simeq 6 \times 10^{58}$~erg for the \AGN~run ($E_{\rm b,noA}\simeq 6 \times 10^{59}$~erg for the \noAGN~run).
Hence the energy from SNe, in the \noAGN~case is comparable to the binding energy of the bulge but because a non-negligible fraction of the SNe energy is lost through radiative cooling, SNe cannot remove the gas from the deep gravitational potential well.
AGN feedback on the other hand delivers an amount of energy larger by more than one order of magnitude than the total binding energy of the bulge, and cleans out its gas content.
Fig.~\ref{fig:lumvsz} shows that the AGN feedback luminosity $L_{\rm AGN}=\dot E_{\rm AGN}$ after $z=9$ is always larger than the feedback luminosity from type II SNe $L_{\rm SNII}=SFR \times \eta_{\rm SN} (10^{51} \rm erg/ 10 M_{\odot})$.
Also the luminosity of the AGN needs to reach high values ($L_{\rm AGN}=10^{11}\, \rm L_\odot$ at $z=7.6$) before it starts to self-regulate the BH growth: this is the point at which the AGN finally unbinds the gas within the bulge.

The final BH mass at $z=6$ is $M_{\rm BH}=8 \times 10^7 \, \rm M_\odot$, somewhat low compared to the brightest quasars observed at $z=6$~\citep{fanetal06}.
Although it is not the aim of this paper to match these observations (we are interested in the early evolution of the progenitor of a typical supermassive cluster), we note that the mass of the halo probed in that simulation turned out to be probably below the mass needed to host these exceptional BHs if ``canonical'' AGN feedback is taken into account.
First, the typical SFRs observed in the far-infrared suggest that the brightest quasars produce stars at $10^3\, \rm M_\odot\,yr^{-1}$, somewhat higher than even the star formation in our \nofeed~simulation (SFR$\simeq 100-200 \, \rm M_\odot\,yr^{-1}$).
Secondly, the maximum halo masses obtained for standard $\Lambda$CDM and for which the SFR is potentially the highest are of the order of several $10^{12}\, \rm M_\odot$~\footnote{The maximum halo mass at $z=6$ in the MassiveBH simulation from~\cite{dimatteoetal12} is $M_{\rm vir}=5 \times 10^{12}\, \rm M_\odot$. In addition, they find a BH mass of $M_{\rm BH}\simeq 2 \times 10^7 \, \rm M_\odot$ for a typical average halo mass of $M_{\rm vir}=5 \times 10^{11}\, \rm M_\odot$ (see figure 1 of~\citealp{degrafetal12}) which is the halo mass simulated here.}.
The number density of observed quasars at $z=6$ is extremely low with one detectable object per comoving $\rm Gpc^3$~\citep{fanetal01}, which suggests that only the most extreme and rarest environments can be the hosts of the brightest quasars in the Universe.
If we depart  from the canonical model and choose a lower AGN efficiency, the BH has to grow to a larger mass in order to release the same amount of energy to reach self-regulation (as already demonstrated in~\cite{duboisetal12agnmodel} using a set of large cosmological simulation boxes with multiple halo masses).
Our test run with an AGN feedback efficiency $\epsilon_{\rm f}$ reduced by a factor of 15 results in a more massive BH of mass $M_{\rm BH}=5\times 10^8 \, \rm M_\odot$ at $z=6.7$.
Note that this reduced efficiency does not significantly change the impact of the AGN feedback on the SFR of the central galaxy, because similar energy is released by the central BH (see Appendix~\ref{appendix:efficiency}).
The BH mass is a factor four below the local $M_{\rm BH}$-$\sigma_{\rm b}$ relationship~\citep{tremaineetal02}, and a factor twelve above the local $M_{\rm BH}$-$M_{\rm b}$ relationship~\citep{haring&rix04} for our canonical model of AGN feedback.
Observations from distant galaxies suggest that such an evolving relationship $M_{\rm BH}$-$M_{\rm b}$ (positive evolution with redshift) is present~\citep{decarlietal10, merlonietal10}, also confirmed by cosmological simulations~\citep{dimatteoetal08, booth&schaye11, duboisetal12agnmodel}.

\begin{figure*}
  \centering{\resizebox*{!}{8.0cm}{\includegraphics{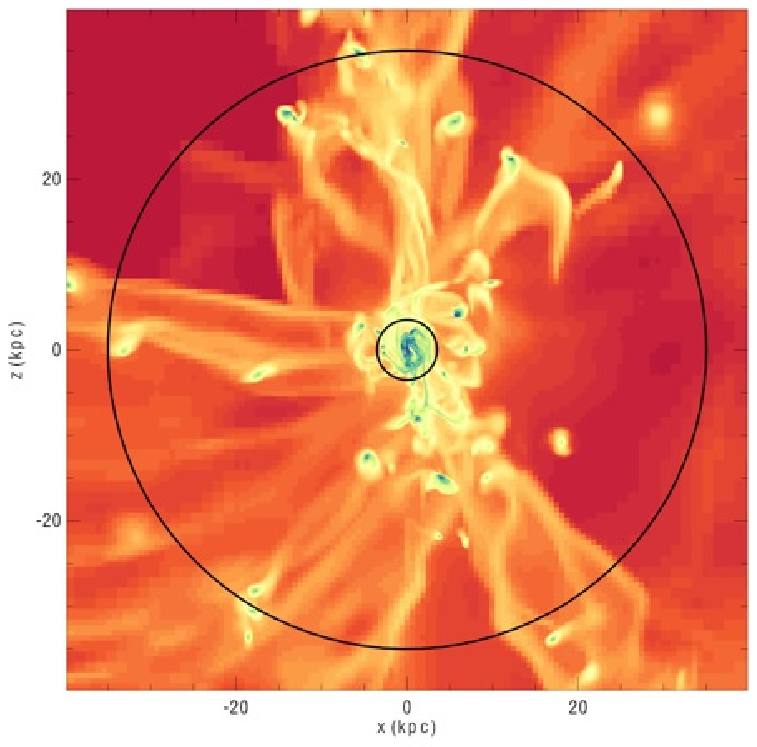}}}\hspace{-0.5cm}
  \centering{\resizebox*{!}{8.0cm}{\includegraphics{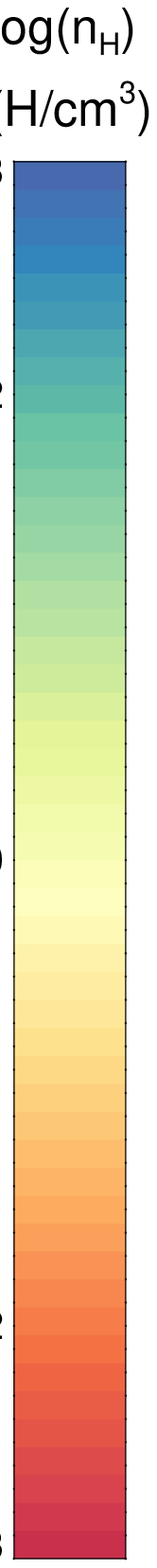}}}
  \centering{\resizebox*{!}{8.0cm}{\includegraphics{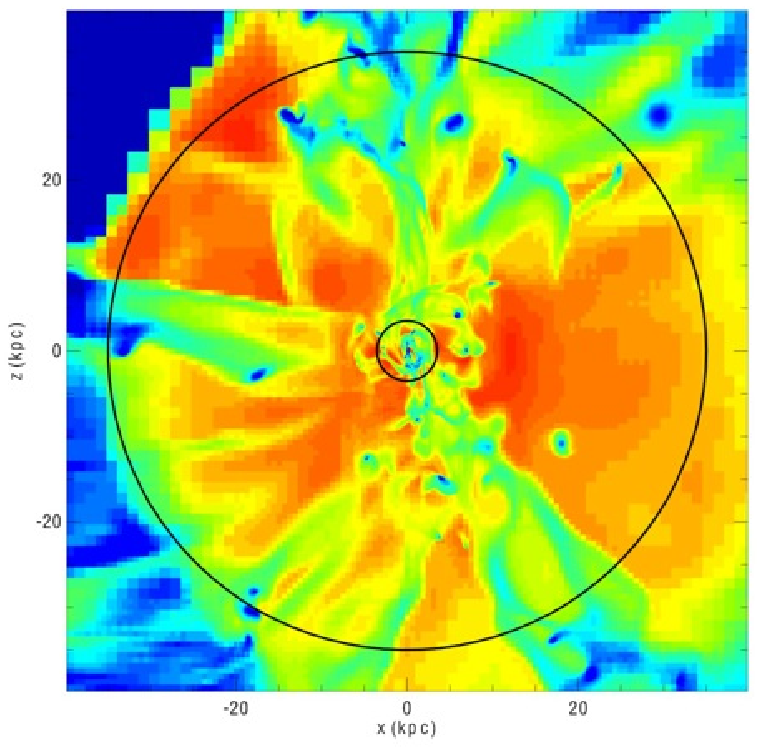}}}\hspace{-0.5cm}
  \centering{\resizebox*{!}{8.0cm}{\includegraphics{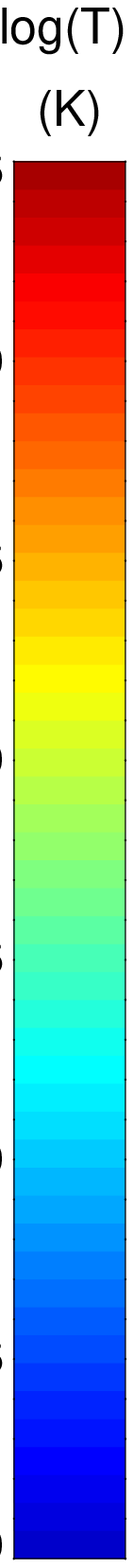}}}\vspace{-1.cm}\\
  \centering{\resizebox*{!}{8.0cm}{\includegraphics{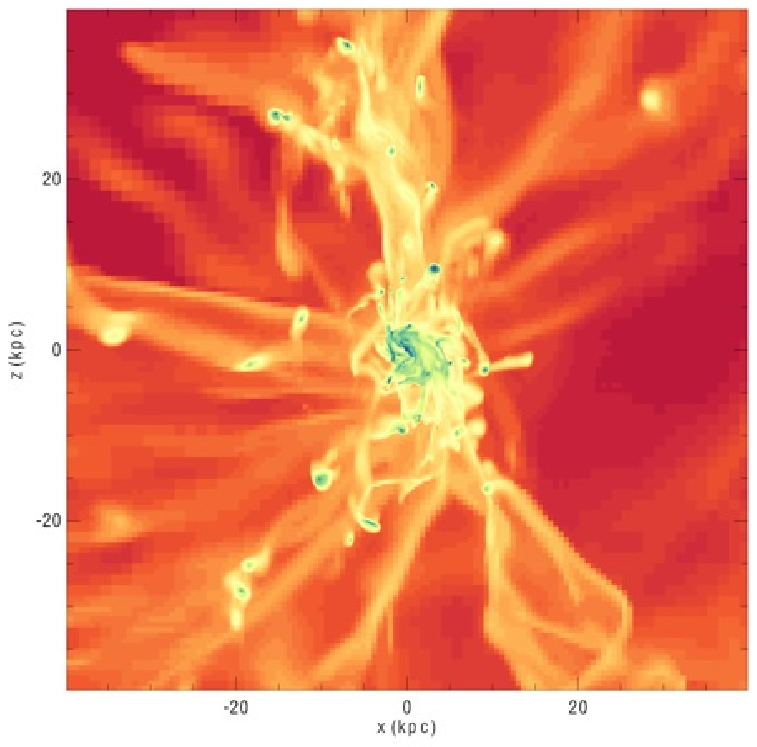}}}\hspace{-0.5cm}
  \centering{\resizebox*{!}{8.0cm}{\includegraphics{./fig/colorbar_density_fig7.ps}}}
  \centering{\resizebox*{!}{8.0cm}{\includegraphics{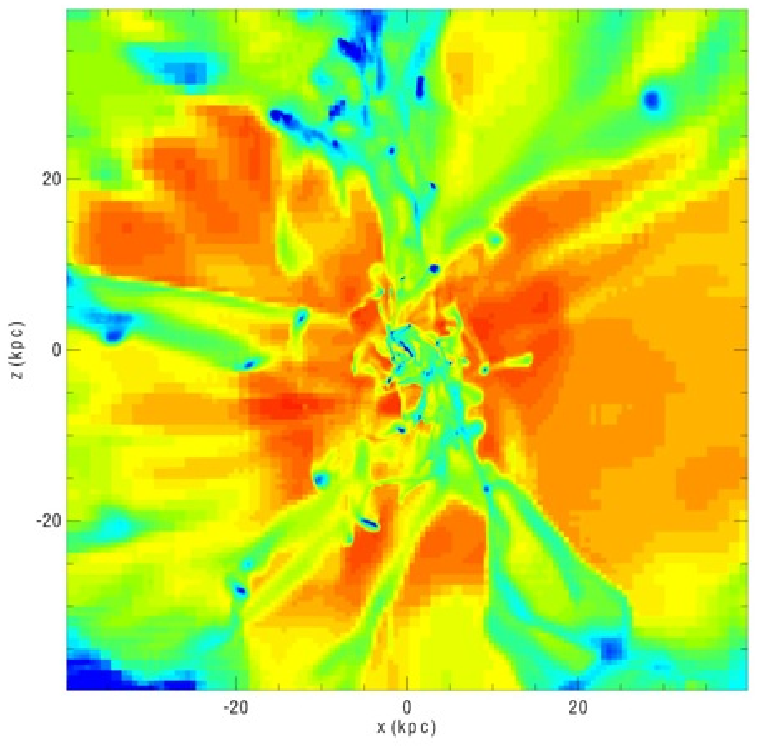}}}\hspace{-0.5cm}
  \centering{\resizebox*{!}{8.0cm}{\includegraphics{./fig/colorbar_temperature_fig7.ps}}}\vspace{-1.cm}\\
  \centering{\resizebox*{!}{8.0cm}{\includegraphics{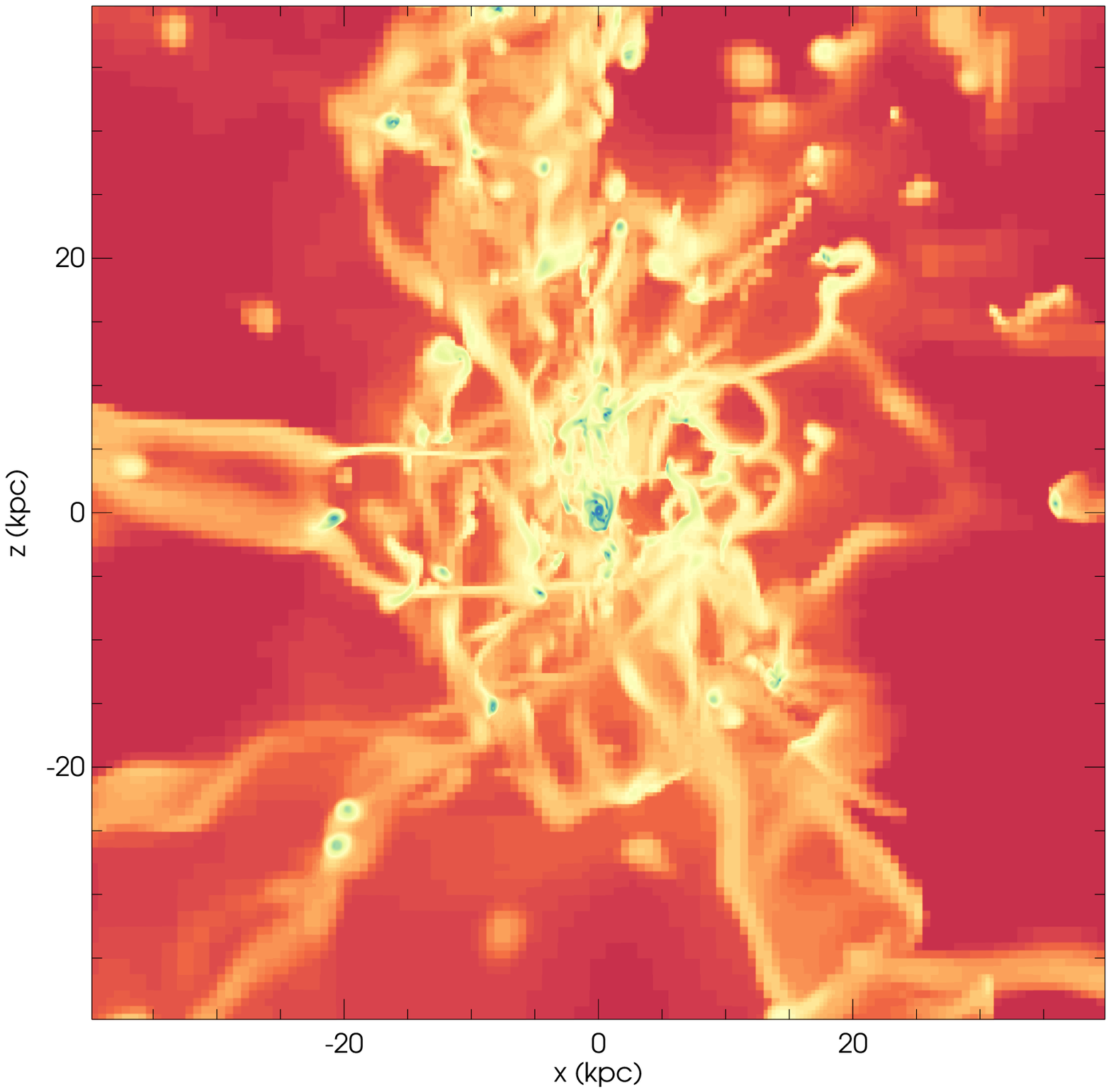}}}\hspace{-0.5cm}
  \centering{\resizebox*{!}{8.0cm}{\includegraphics{./fig/colorbar_density_fig7.ps}}}
  \centering{\resizebox*{!}{8.0cm}{\includegraphics{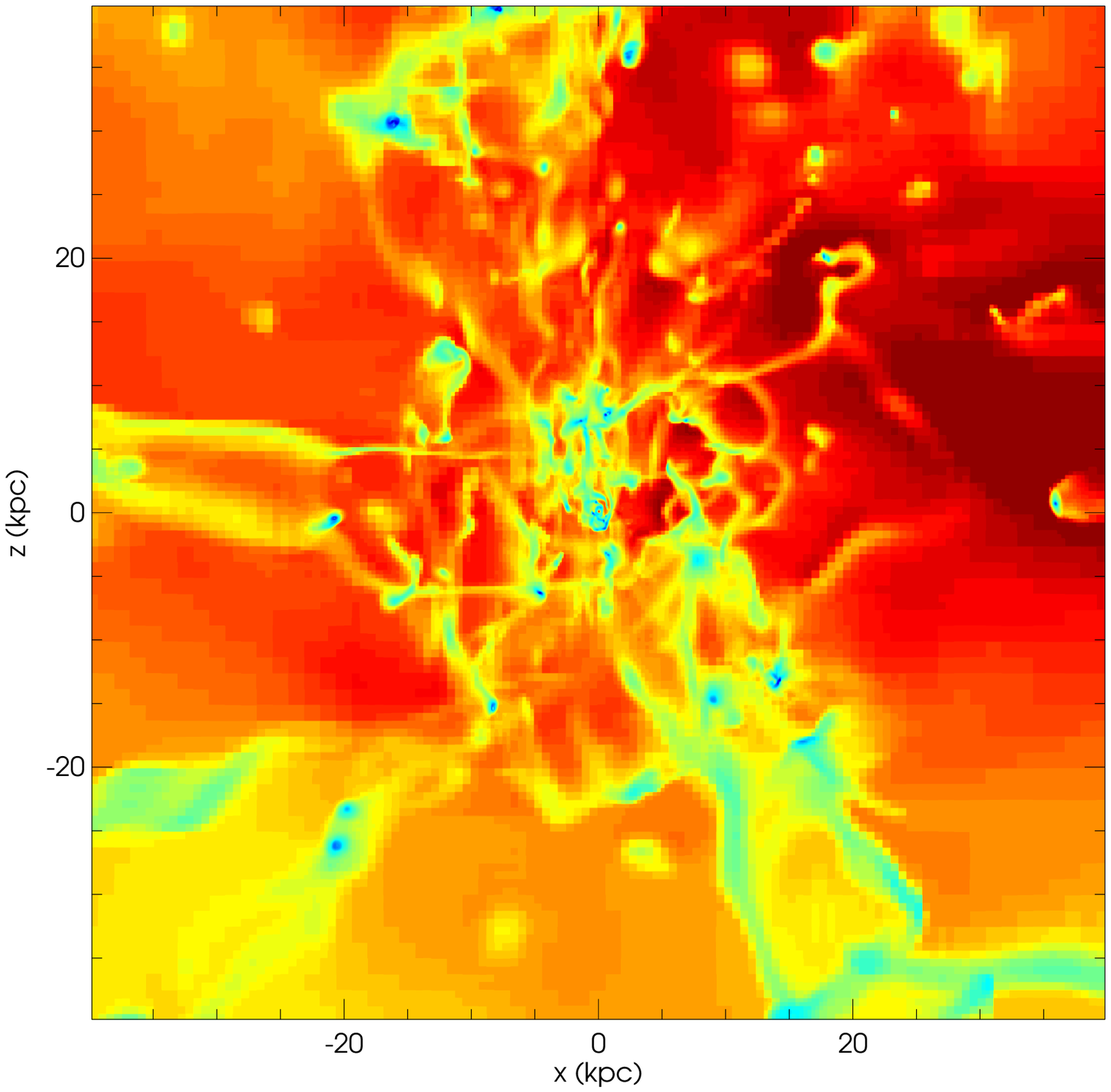}}}\hspace{-0.5cm}
  \centering{\resizebox*{!}{8.0cm}{\includegraphics{./fig/colorbar_temperature_fig7.ps}}}
  \caption{Mass-weighted projected number densities (left panels), and temperatures (right panels), for the \nofeed~case (top panels), the \noAGN~case (middle panels), and the \AGN~case (bottom panels) at $z=6$. The projections are 80 kpc deep. The small and large circles are respectively $0.1\, r_{\rm vir}$ and $r_{\rm vir}$. Both simulations without AGN feedback show strikingly similar features with cold streams of gas connecting to the central galaxy in a continuous manner. In the \AGN~case, these streams are strongly disturbed and the gas temperature is increased and expands to larger distance. }
    \label{fig:verynice_dT}
\end{figure*}

\section{Removing gas from the halo}
\label{section:outflow}

AGN feedback has an important impact on the  mass content within the galaxy, and also on larger scales by virtue of it removing gas efficiently from the halo.
Fig.~\ref{fig:verynice_dT} shows the gas density and gas temperature distribution in the halo at $z=6$ for the three different runs.
From a comparison of the \nofeed~(upper panels) and \noAGN~(middle panels) cases, it appears that SN explosions in the \noAGN~run do not perturb the cold gas streaming toward the central galaxy: the same features are present as in the \nofeed~case, and the cold gas connects all the way down to the central galaxy.
The most obvious difference between the two simulations is that the hot gas filling the halo is slightly more extended in the simulation with SNe explosions.
The difference with the \AGN~case however is striking: very hot (T$\sim10^7$~K) gas pours into the circum-galactic medium and the dense cold streams are morphologically perturbed within the halo.
We will now quantify the extent to which AGN feedback is able to remove gas from the halo.

\begin{figure}
  \centering{\resizebox*{!}{6.cm}{\includegraphics{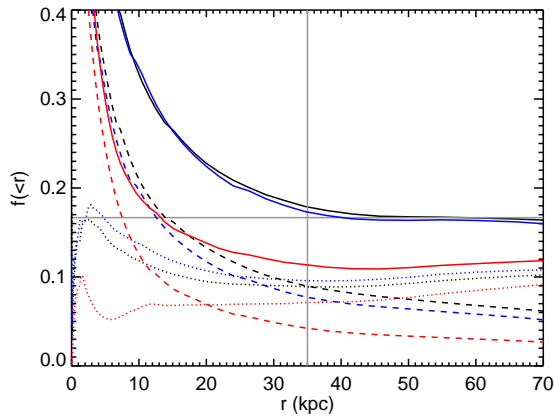}}}
  \caption{Baryon fraction (solid), stellar fractions (dashed), gas fractions (dotted) at $z=6$ for the \nofeed~case (black), \noAGN~case (blue), and for the \AGN~case (red). The horizontal grey line is the universal baryon fraction. Baryon fractions in the \nofeed/\noAGN~cases at $r_{\rm vir}$ (vertical grey line) are at the universal baryon fraction. The AGN feedback significantly reduces the baryon fraction at $r_{\rm vir}$ by 30 per cent below its universal value.}
    \label{fig:fractionvsr}
\end{figure}

\begin{figure}
  \centering{\resizebox*{!}{6.5cm}{\includegraphics{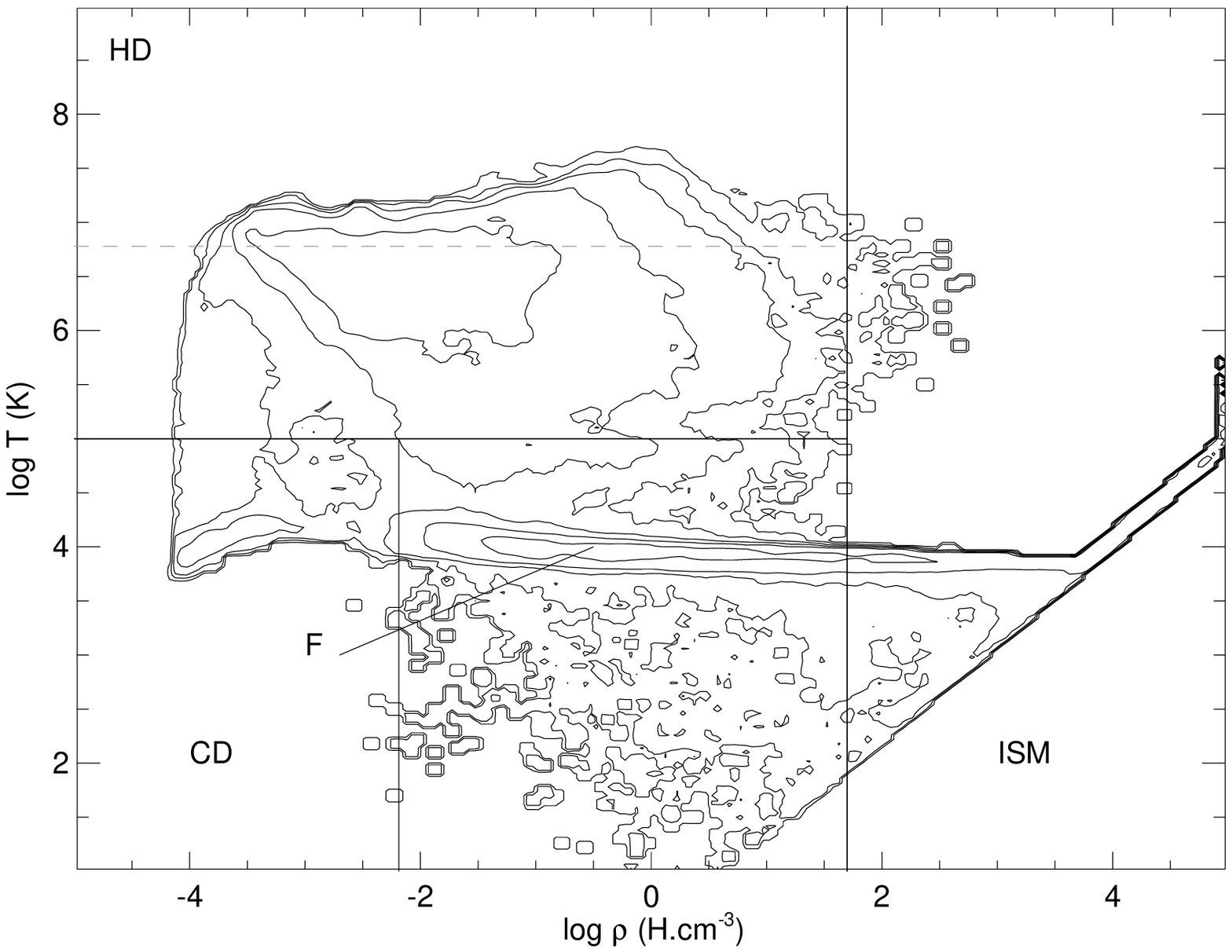}}}\vspace{-0.5cm}
  \centering{\resizebox*{!}{6.5cm}{\includegraphics{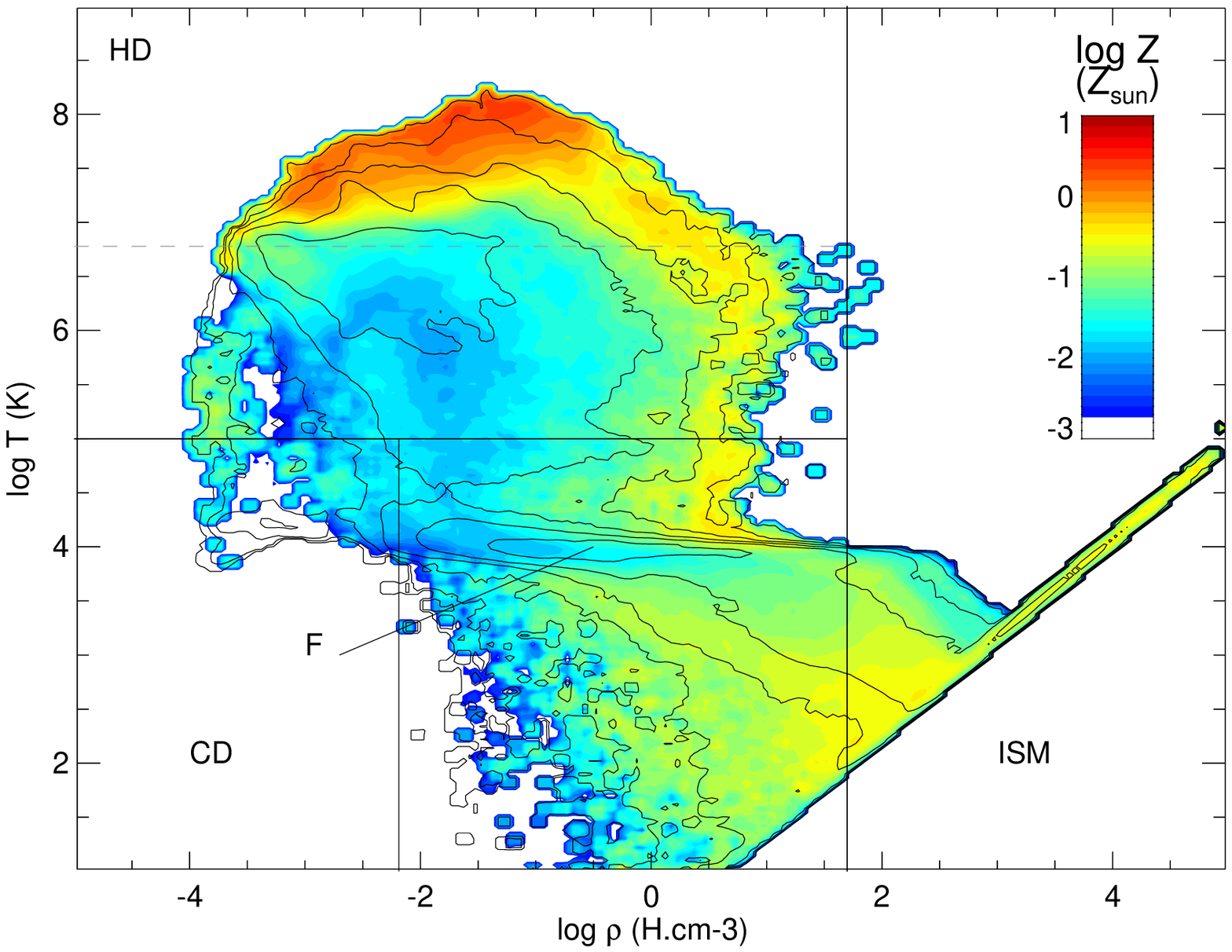}}}\vspace{-0.5cm}
  \centering{\resizebox*{!}{6.5cm}{\includegraphics{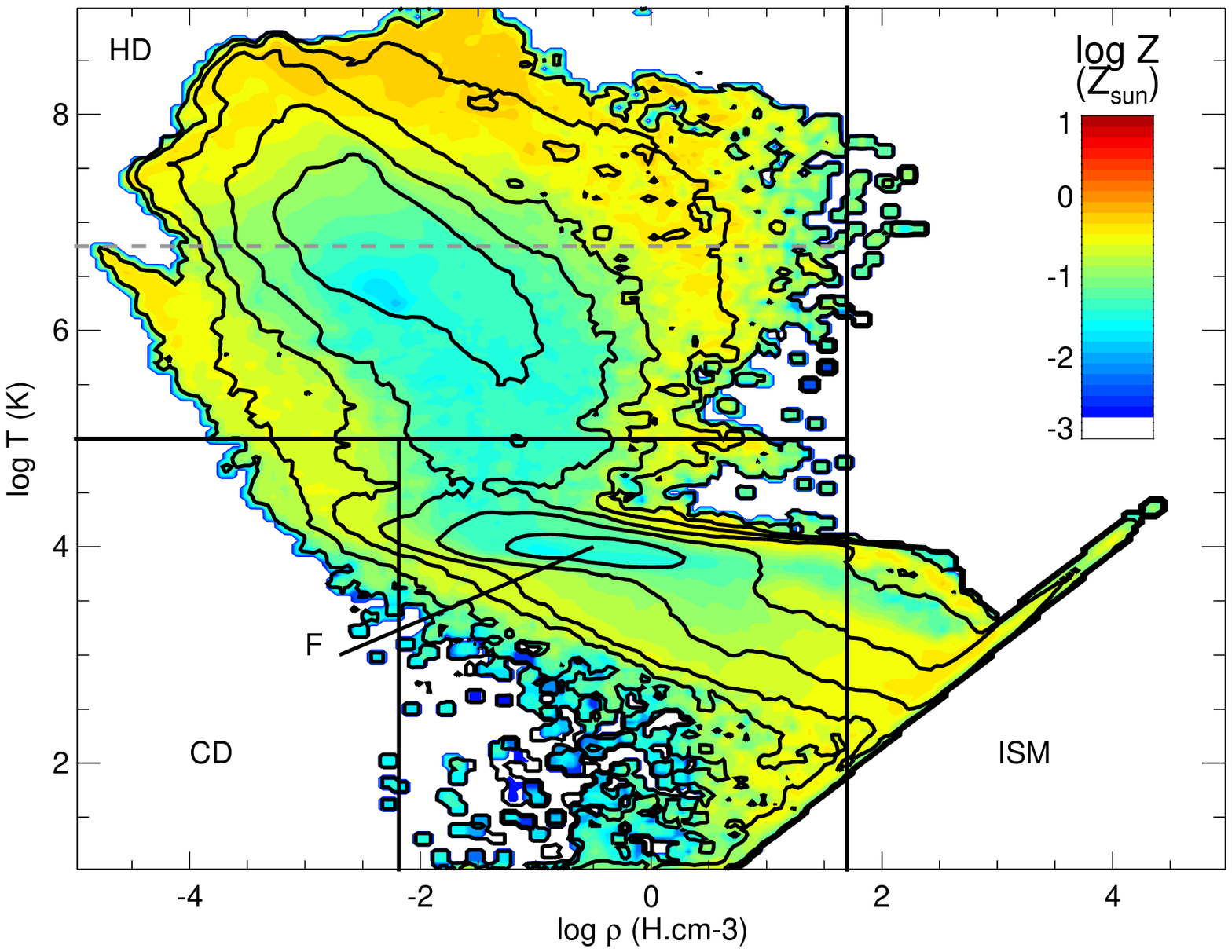}}}
  \caption{Temperature-density diagrams of all the gas within $r_{\rm vir}$ for the \nofeed~case (top), \noAGN~case (middle) and the \AGN~case (bottom) at $z=6$. Colors represent the logarithm of the metallicity of the gas while the black solid contours are the mass-weighted contours with 1 dex difference between each contour. The solid horizontal and vertical lines separate the four different phases of the gas (see text for details). The dashed horizontal line is the virial temperature of the halo at $z=6$. }
    \label{fig:histo}
\end{figure}

Fig.~\ref{fig:fractionvsr} displays the baryon fraction as a function of radius for the three simulations at $z=6$.
While we see that SN feedback alone has almost no effect on the total baryon fraction at the virial radius $r_{\rm vir}$, the addition of AGN feedback drives more than 30 per cent of the baryons out of the halo (beyond two $2\, r_{\rm vir}$ that is the size of the high resolution region) and decreases the total stellar fraction (or mass) by a factor 2.
This confirms that strong and early feedback activity can significantly reduce the total amount of baryons in the halo.
Feedback at later times, might be less efficient because halos tend to be more spherical, making it more difficult for gas to escape from collapsed structures~\citep{peiranietal12}.
We also demonstrate that a standard recipe for SN feedback does not suffice to alter the baryonic mass content in a progenitor of a massive halo, whereas AGNs, with their large energy release (more than a factor 10 compared to the energy released by SNe), are good candidates for powerful early pre-heating.

It is important to investigate the ability of AGN feedback to efficiently remove the gas from the halo, and whether the different gas phases are able to survive within the hot atmosphere produced by ejecta from the central BH.
We devide the gas into four different phases: the \emph{star forming} gas (ISM) with gas density above $\rho \geÊ\rho_0$, the \emph{collimated cold flows} (F) with gas density $100 \rho_{\rm avg} \le \rho < \rho_0$ and temperature $T< 10^5$ K, the \emph{cold diffuse} gas (CD) with $\rho < 100Ê\rho_{\rm avg}$ and $T<10^5$ K, and the \emph{hot diffuse} gas (HD) with $\rho < \rho_0$ and $T\ge 10^5$ K, where we define $\rho_{\rm ave}$ as the cosmic average gas density.
Fig.~\ref{fig:histo} shows temperature-density diagrams of the gas metallicity and of the gas mass distribution of the gas within the virial radius of the halo at $z=6$.
Even without SNe or AGN feedback, a hot phase with a temperature close to the virial temperature of the halo, $T_{\rm vir}=6\times 10^6$~K is present.
This hot phase originates from cosmic gas that shock-heats as it  accretes onto the halo, and for which the radiative cooling is not efficient enough~\citep{birnboim&dekel03}.
When SNe deliver thermal energy to the gas, the temperature of the hot gas component rises above the virial temperature of the halo.
AGN feedback increases the temperature of the hot gas even further: the mass-weighted temperature of the hot gas phase is now $\bar T_{\rm HD, AGN}=8\times 10^6$~K instead of  $\bar T_{\rm HD, noAGN}=3.5\times 10^6$~K (similar to the \nofeed~case).
The bursts of AGN feedback also produce more mixing of metals in the hot diffuse gas, as well as in the cold diffuse gas: we will confirm this effect below by analyzing the morphological structure of the filaments.
Pushing the metal rich gas out of the halo centre and increasing its mixing are aspects of AGN feedback believed to play important roles in getting realistic flat metallicity profiles at late times in galaxy clusters~\citep{leccardi&molendi08, fabjanetal10, mccarthyetal10}
 
\begin{figure*}
  \centering{\resizebox*{!}{6.5cm}{\includegraphics{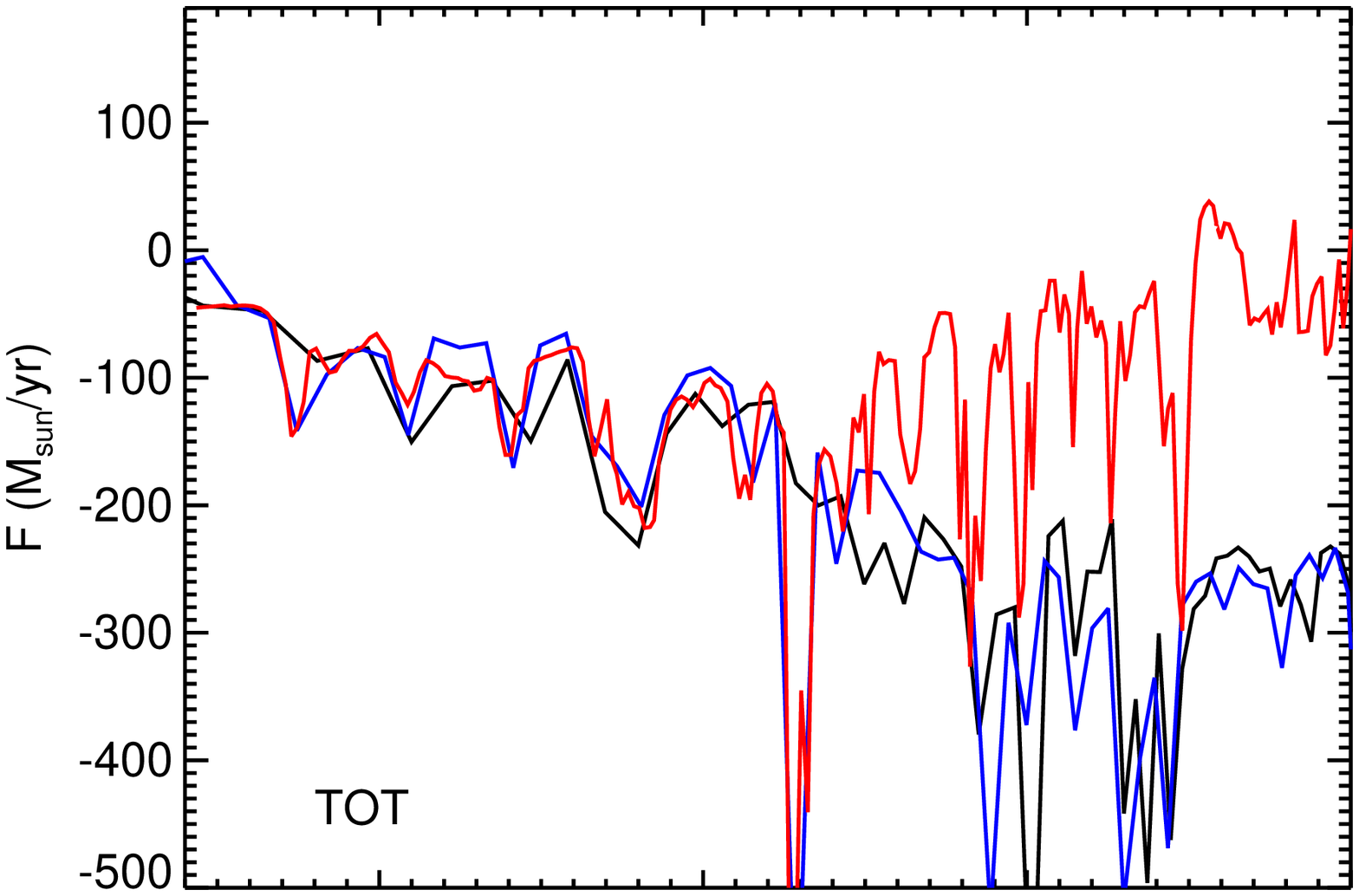}}}\hspace{-1.75cm}
  \centering{\resizebox*{!}{6.5cm}{\includegraphics{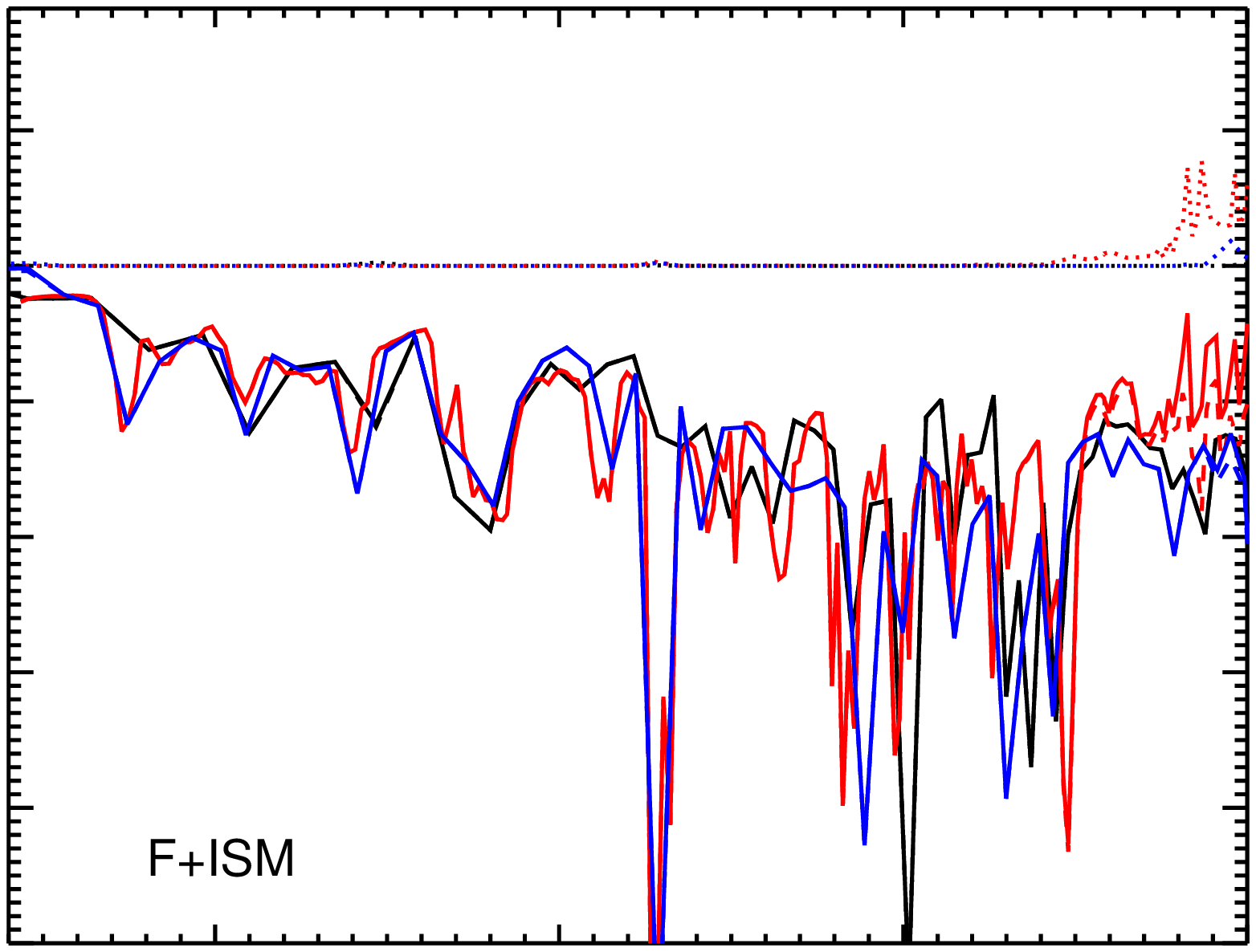}}}\vspace{-1.3cm}
  \centering{\resizebox*{!}{6.5cm}{\includegraphics{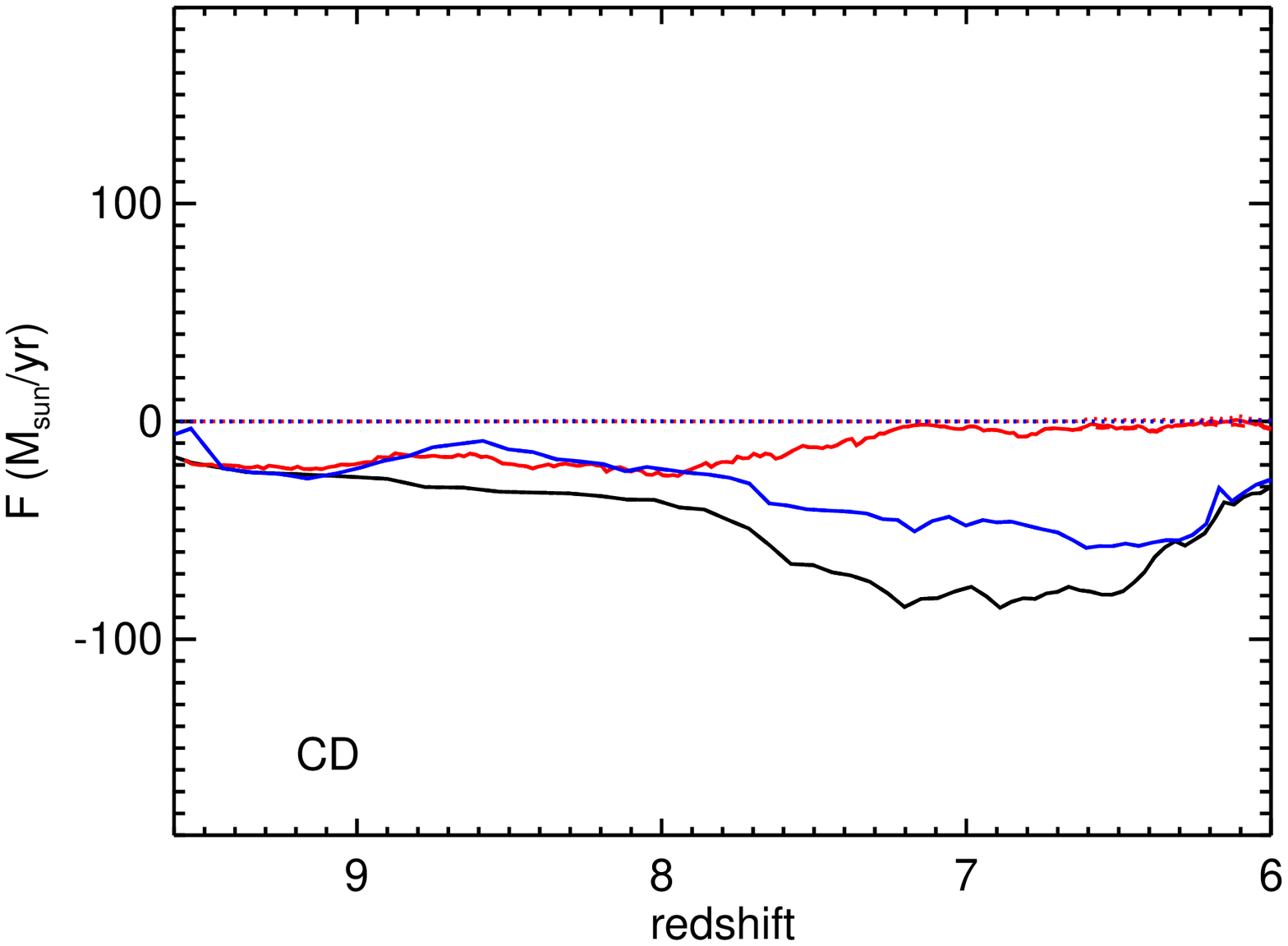}}}\hspace{-1.75cm}
  \centering{\resizebox*{!}{6.5cm}{\includegraphics{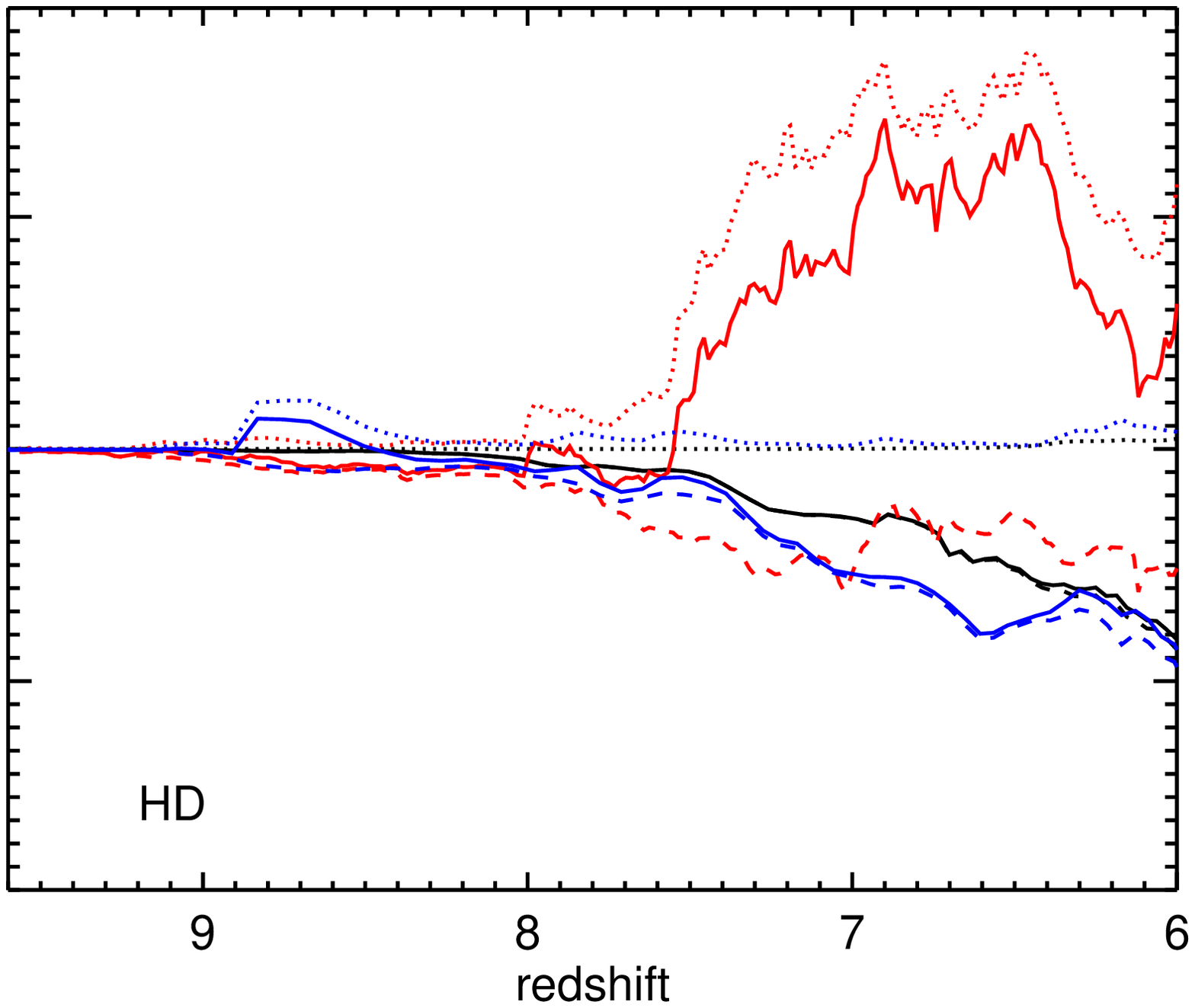}}}
  \caption{Mass fluxes measured at $r_{\rm vir}$ as a function of redshift for the \nofeed~case (black) \noAGN~case (blue), and for the \AGN~case (red). Net fluxes are solid lines, outflow components are dotted lines, and inflow components are dashed lines. Top left panel is the total flux of gas, top right panel are the cold flows and star forming gas, bottom left panel is the cold diffuse only, and bottom right is the hot phase only. Bottom panels have a different scaling of the y-axis than the top panels. Global accretion is significantly reduced due to AGN feedback, CD accretion is quenched, HD accretion is strongly reduced with a strong outflowing component, and the F$+$ISM component is slightly reduced with some cold gas being transported outside of the halo.}
    \label{fig:fxnrvir}
\end{figure*}

Fig.~\ref{fig:fxnrvir} shows the mass fluxes measured at $r_{\rm vir}$ for the \noAGN~and \AGN~runs as a function of redshift.
Outflow and inflow components (positive and negative fluxes) are separated according to their radial velocities: positive radial velocities correspond to outflows and vice versa.
The \noAGN~case reveals that the majority of the mass is accreted in the form of cold flows, with an increasing amount of inflowing diffuse (hot and cold) gas as the halo becomes sufficiently massive to enter a phase where both cold flows and shock-heated gas coexist~\citep{ocvirketal08}.
In the \AGN~run, cold diffuse accretion is quenched, hot diffuse accretion is significantly reduced, cold accretion is slightly reduced, and large amounts of hot gas are driven beyond the virial radius ($>100\,\rm M_\odot\,yr^{-1}$).
The hot outflowing phase manages to reduce the net flux to less than $<100\,\rm M_\odot\,yr^{-1}$ entering the halo at $z=6$ as opposed to a flux of $\sim 250 \,\rm M_\odot\,yr^{-1}$ without AGN.
The intensity of the mass outflow generated by the central AGN is comparable to the SFR measured in the absence of feedback ($\sim 100 \, \rm M_\odot\,yr^{-1}$, see Fig.~\ref{fig:sfr}), thus, the mass loading factor of the wind is large enough to self-regulate the SFR inside the galaxy.
The outflow velocity of the hot gas at $r_{\rm vir}$ is $\sim 500\, \rm km\, s^{-1}$ averaged over time during the removal of the hot gas from the halo, with values close to the galaxy that can reach several $1000\, \rm km\, s^{-1}$ consistent with outflow velocities observed in powerful distant quasars~\citep{maiolinoetal05,maiolinoetal12, cano-diazetal12}.
This value is larger than the escape velocity of the halo $v_{\rm esc}=350\, \rm km\, s^{-1}$ at $z=6$ and explains why the baryon fraction within the halo is efficiently reduced. 
Note that the outflow rate of the hot diffuse phase does not continuously increase but instead endures a sharp drop after $z=6.5$ corresponding to the global decline of the central quasar activity after it enters its sub-Eddington phase (at $z=7.5$, see Fig.~\ref{fig:lumvsz}) with a time lag.
At late times, around $z=6$, some of the cold filamentary gas is entrained by the hot gas and flows out beyond $r_{\rm vir}$, but most of the gas continues to be accreted.

Fig.~\ref{fig:mass4com} shows the integrated mass of star forming gas, cold filaments and hot gas at $z=6$ for the three different simulations. 
SN feedback is not able to remove the cold star forming ISM from the central galaxy, and the filamentary distribution is not affected by these explosions.
Only the hot gas is slightly removed from the central region of the galaxy by SNe.
AGN feedback removes dense gas from the galaxy and pushes hot gas to very large distances.
Furthermore the amount of cold filamentary material in the core of the halo (at $0.1\, r_{\rm vir}$) is decreased by one order of magnitude, suggesting that filaments are also pushed away to larger distances.
This is  confirmed by Fig.~\ref{fig:fxnrvir} showing an outflow of cold filaments at $r_{\rm vir}$ around $z=6$.
The removal of cold streams from the centre of the halo can potentially reduce the Lyman-$\alpha$ emission of simulated Lyman-$\alpha$ blobs, which usually tend to be brighter than observed~\citep[see][where no AGN feedback is accounted for]{rosdahl&blaizot12}.
However a full treatment of the radiative transfer from stars and the central AGN is required to track this emission.

As visual inspection of the gas density in Fig.~\ref{fig:verynice_dT} already suggests, the cold filamentary structures exhibit morphological differences between the \nofeed/\noAGN~cases and the \AGN~case.
We quantify these differences using the skeleton~\citep{sousbieetal09}, a geometric 3D ridge extractor well suited to identify filaments.
We build the skeleton of the gas density within the virial radius of the halo for the \noAGN~and the \AGN~cases at different redshifts, and compare the properties of the skeleton such as its average length and curvature.
The gas density is sampled over a uniform grid of $256^3$ pixels inside a box of 1.1 comoving Mpc ($4.5\, r_{\rm vir}=160$ physical kpc at $z=6$) and is smoothed with a Gaussian filter of  {$4\sigma$}.
Once the skeleton is computed, it is thresholded both within $r_{\rm vir}$ and where gas number density is larger than $n_{\rm H}> 50 \, \rm H\, cm^{-3}$.
The two skeletons are globally very similar but depart significantly in shape on small scales (Fig.~\ref{fig:skel_length}), with the  \AGN~skeleton being both shorter and curlier with decreasing redshift relative to the \noAGN~skeleton. 
More specifically, Fig.~\ref{fig:skel_length} shows the ratio between the \AGN~and the \noAGN~case of the total length of the thresholded skeletons as well as the ratio of their mean curvature.
As expected, compared to the \noAGN~case, the total length of the skeleton in the \AGN~case decreases by about 50 per cent, while its curvature \citep[as defined in][]{pogosyanetal09} increases by about 20 per cent.
These are clear signatures that AGN feedback is both destroying the connectivity of the filaments and bending them.

\section{Conclusion and Discussion}
\label{section:conclusion}

Using high resolution cosmological hydrodynamical simulations to follow the formation of the progenitor of a BCG at $z=6$, we have studied the impact of early BH growth and its accompanying AGN activity on the formation of its host proto-galaxy and on its environment. Our BH and AGN model is calibrated to match the BH-host relationships at z=0.
The main results of this investigation are the following:

\begin{figure}
  \centering{\resizebox*{!}{6cm}{\includegraphics{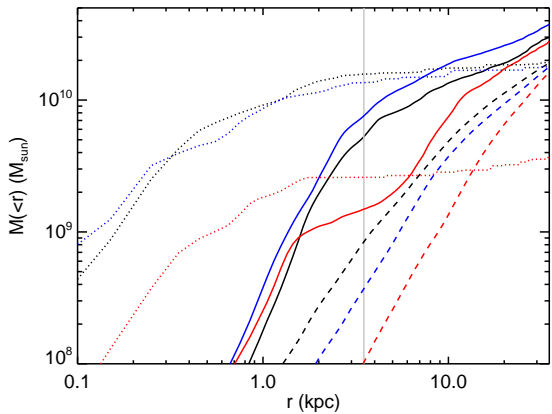}}}
  \caption{Integrated mass profiles at $z=6$ of the cold filaments (solid), star forming gas (dotted), and hot gas (dashed) for the \nofeed~(black), \noAGN~(blue), and \AGN~case (red). The vertical grey line corresponds to $0.1\,r_{\rm vir}$. Profiles in the \noAGN~case are very similar to the \nofeed~case: only a small quantity of hot gas is removed by SNe. AGN feedback however expels large amounts of gas from the centre of the halo. }
    \label{fig:mass4com}
\end{figure}

\begin{figure}
  \centering{\resizebox*{!}{6.5cm}{\includegraphics{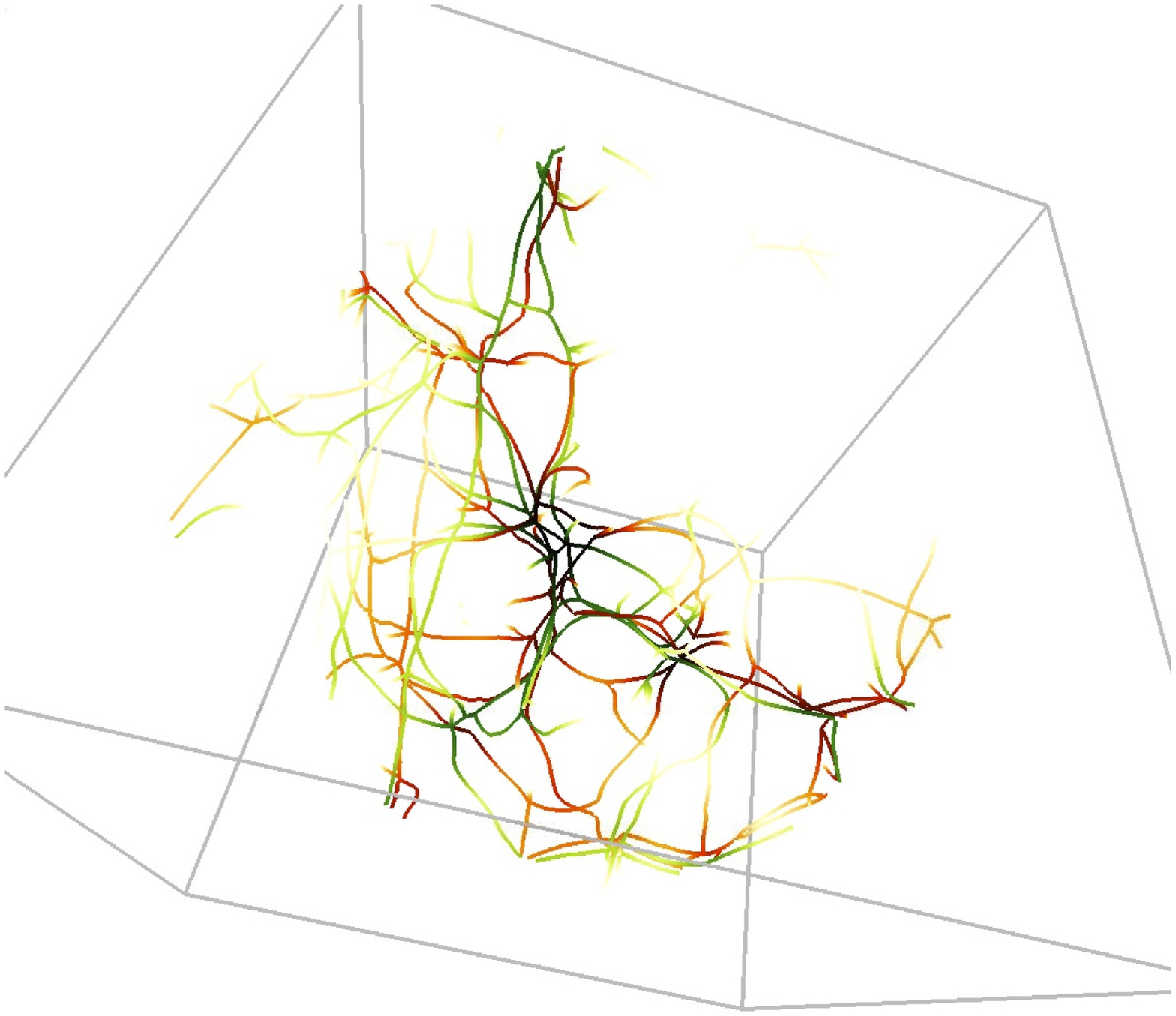}}}
  \centering{\resizebox*{!}{5.5cm}{\includegraphics{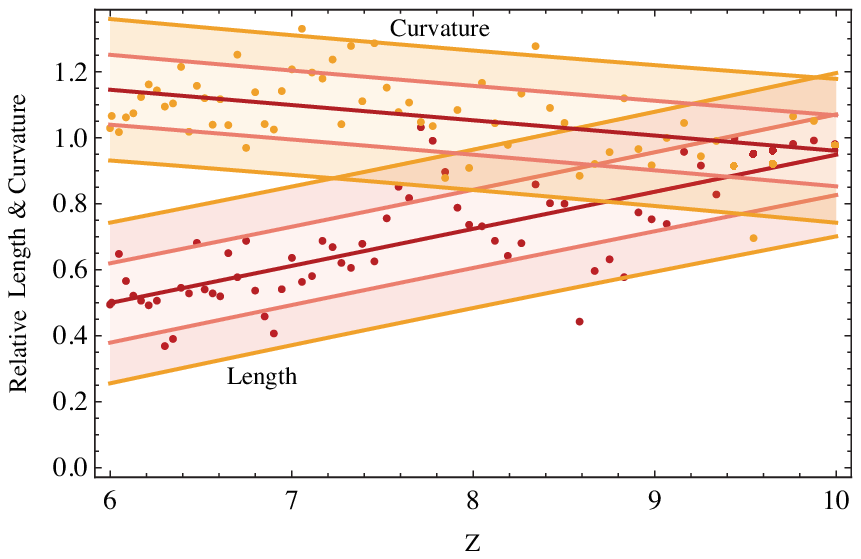}}}
  \caption{\emph{Top:} Image of the skeleton for the \noAGN~case (from black to green) and the \AGN~case (from black to orange). \emph{Bottom:} The ratio of the \AGN~to the \noAGN~length (bottom bundle, red points) and curvature (top bundle, orange points) of the skeleton within the halo as a function of redshift (shaded areas are 1 and 2 $\sigma$ deviations). AGN feedback changes the structure of the filamentary network by bending and shortening it.}
    \label{fig:skel_length}
\end{figure}

\begin{itemize}
\item{SN feedback is extremely inefficient at regulating star formation in progenitors of massive galaxies and has almost no impact on the surrounding gas of the galaxy.}
\item{AGN feedback through a quasar mode quenches the SF in the bulge and reduces the amount of gas accreted onto the central galaxy and available for star formation.}
\item{With $r_{\rm b}= 50$~pc and a stellar mass of $M_{\rm b}=6.2\times 10^9\, \rm M_\odot$, the stellar bulge in our simulation with AGN feedback is still a factor ten smaller than the most compact observed stellar bulges at $1<z<3$ of comparable stellar mass. This suggests that the strong evolution of bulge sizes between $z=0$ and $z=2$ continues to higher redshift.}
\item{The distribution of mass within the core of the halo is strongly affected by the central AGN: gas, stars and DM exhibit shallower density profiles in their centre. The adiabatic contraction of DM in the core of the halo is reduced by 35 per cent due to the presence of the central AGN compared to the simulation without any feedback. }
\item{The cold baryon content is regulated by two stages of BH growth: (1) the rapid early growth of the BH at the Eddington limit driven by direct cold filamentary infall onto the bulge, followed by (2) a sub-Eddington phase where bursts of AGN activity are triggered by inflows and rapid clump migration towards the BH, brought on by merger induced torques.}
\item{The self-regulation of the BH growth due to AGN feedback limits the BH mass to a value a factor four below the $z=0$ $M_{\rm BH}$-$\sigma_{\rm b}$ relation and a factor 12 above the $z=0$ $M_{\rm BH}$-$M_{\rm b}$ relation. Our simulations therefore suggest that these relations evolve with redshift.}
\item{AGN feedback drives large-scale outflows that reduce the baryon fraction in the halo by 30 per cent. The gas that is expelled from the halo is essentially a hot component with temperature well above virial, and with an outflow velocity of $500\, \rm km\, s^{-1}$ at $r_{\rm vir}$ that is larger than the escape velocity of the halo.}
\item{Cold filaments are morphologically disturbed by AGN with a small fraction of them blown out of the halo, and a deficit of them in the core of the halo close to the galaxy at $0.1\, r_{\rm vir}$.}
\end{itemize}

For ``canonical'' feedback parameters (see section~\ref{section:numerics}), we demonstrated the key role played by AGN feedback in regulating both the star formation, and the amount of gas available in the halo for the progenitor of a massive cluster of galaxies.
We also tested two variations of our canonical model for AGN feedback with the following results.
A model where AGN feedback energy is only deposited in a \emph{radio} (jet) mode produces bipolar outflows similar to those  produced by our canonical model in which energy is released in isotropic thermal bursts;  thus, the impact on the surrounding gas is as efficient as in the canonical case (see Appendix~\ref{appendix:jet}).
Reducing the AGN feedback efficiency of our canonical model by one order of magnitude also does not change the AGN's impact on the star formation in the central galaxy: the BH grows to larger masses so as to  release the same amount of energy to unbind the bulge gas (see Appendix~\ref{appendix:efficiency}).
These very simple tests lend support to our results on the impact of AGN feedback, showing them to be robust to  variations of its modeling.
In the future, alternative variations could also be explored: for instance, gravitational recoil of BHs which commonly takes place in massive halos~\citep{volonterietal10}; a mode of quasar heating where the radiative transfer of photons and their radiation pressure is followed self-consistently~\citep{kimetal11}.

Our numerical experiments have possible implications for models of pre-heating of the gas at high redshift.
Standard pre-heating models suppose that gas elements are more or less uniformly heated (either by imposing a temperature or an entropy floor for the gas) to some value  large enough to prevent the formation of excessively luminous clusters and to break the self-similarity of clusters~\citep{bialeketal01, babuletal02, borganietal02, muanwongetal02, voitetal03}.
However, these models fail to predict some of the basic properties of clusters, producing large isentropic cores~\citep{borganietal05, younger&bryan07} and shallow pressure profiles~\citep{kayetal12}.
Also the uniform pre-heating leaves the IGM with the wrong covering fraction for the Lyman-$\alpha$ forest unless only high gas density regions are pre-heated~\citep{shangetal07, borganietal09}.
Our simulations demonstrate that AGN feedback is a potential strong source of pre-heating, but instead of \emph{uniformly} changing the specific energy of the gas, it changes the temperature, and thus the entropy of the \emph{hot diffuse gas} component, leaving the thermodynamical (but not dynamical) properties of the cold streams unchanged.
Thus, one can get a more realistic model of pre-heating by only changing the temperature of the gas that is already hot (above $10^5$~K).

There are several potential worries concerning our results and the probable consequences on galaxy formation at lower redshift.
The SFR measured in the central galaxy at $z=6$ is only $20 \, \rm M_\odot\,yr^{-1}$. A naive back-of-the envelope calculation predicts that, with this amount of SFR, the final BCG mass would be of the order of $10^{11} \, \rm M_\odot$, one order of magnitude below the expected mass for BCGs ($10^{12} \, \rm M_\odot$,~\citealp{mosteretal10}).
However, the central galaxy can build a significant fraction of its mass through merger events.
With the SFRs of our AGN feedback case, more than 90 per cent of the stellar mass would need to be added by ex-situ star formation (as opposed to in-situ). Hydro simulations for the most massive structures show that this is plausible~\citep{oseretal10}.
Our result is in agreement with observations of $z=6$ galaxies of comparable stellar mass ($\sim 10^{10}\, \rm M_\odot$) that show a SFR of the same level $20 \, \rm M_\odot\,yr^{-1}$~\citep{starketal09}, even though taking into account correct dust extinction would increase the SFR by a factor 2~\citep{bouwensetal12}.
Also, the strong outflows and self-regulation of the cold baryons produced by our AGN model provides a suitable solution to suppress the SFR between $4<z<7$ in order to reproduce the plateau of the specific SFR above $z>2$~\citep{weinmannetal11}, and to decrease the specific star formation rate that is required for the most massive galaxies~\citep{kimmetal12}.

Another potential tension is the low baryon fraction obtained in the AGN feedback simulation, i.e. $f_{\rm b}=0.11$ at $z=6$. 
Observed clusters in this mass range $M_{\rm vir}=2\times 10^{15}\,\rm M_\odot$ at $z=0$ exhibit a universal baryon fraction~\citep{gonzalezetal07, giodinietal09}.
However, this discrepancy is an issue only if it is assumed that the gas cannot re-collapse.
Simulations of strong pre-heating of structures show that the gas fraction is reduced early on, and is followed by re-accretion of the hot expelled gas that raises up the baryon fraction at lower redshift~\citep{fgetal11, peiranietal12}. 
The capacity of AGN feedback to expel gas from halos is a key problem for precision cosmology and future surveys, as large-scale simulations have shown that the presence of baryons and their associated physics modify by a few percent the measurements of the matter power spectrum on Mpc scales where possible signature of modified gravity are sought~\citep{guilletetal10, vandaalenetal11}.

In order to put stronger observational constraints on our current model of AGN feedback and confirm or invalidate the plausibility of our theoretical predictions, simulations of more massive halos at $z=6$ have to be carried out (in the spirit of~\citealp{dimatteoetal12}).
Another direction should also be explored, by simulating the formation of intermediate-redshift $z=3-1$ galaxies where the direct imaging of galaxies is able to unveil the clumpiness of these gas-rich objects~\citep{elmegreenetal07}, and put more severe constraints on the AGN-galaxy connection~\citep{bournaudetal12}.

\section*{Acknowledgments}
The simulations presented here were run on the DiRAC facility jointly funded by STFC, the Large Facilities Capital Fund of BIS and the University of Oxford. 
This research is part of the Horizon-UK project. 
CP thanks T. Sousbie and C. Gay for skeleton related codes. 
YD and JS acknowledge support by the ERC advanced grant (Dark Matters).

\bibliographystyle{mn2e}
\bibliography{author}

\appendix

\section{Isotropic wind versus collimated jet inputs}
\label{appendix:jet}

\begin{figure}
  \centering{\resizebox*{!}{7.5cm}{\includegraphics{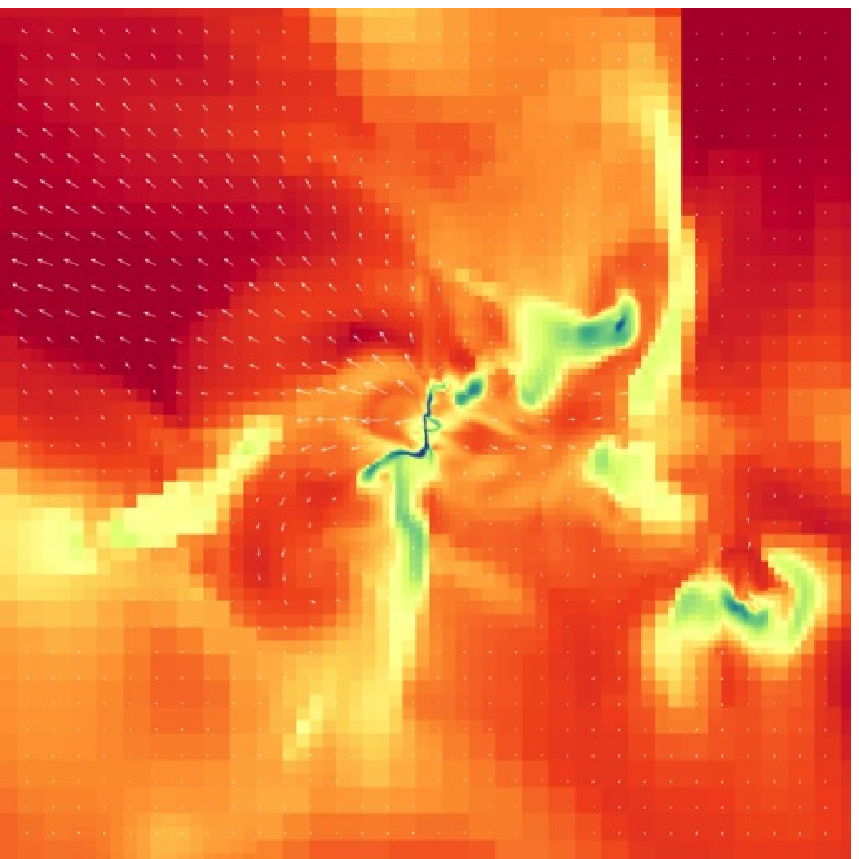}}}\vspace{0.25cm}
  \centering{\resizebox*{!}{7.5cm}{\includegraphics{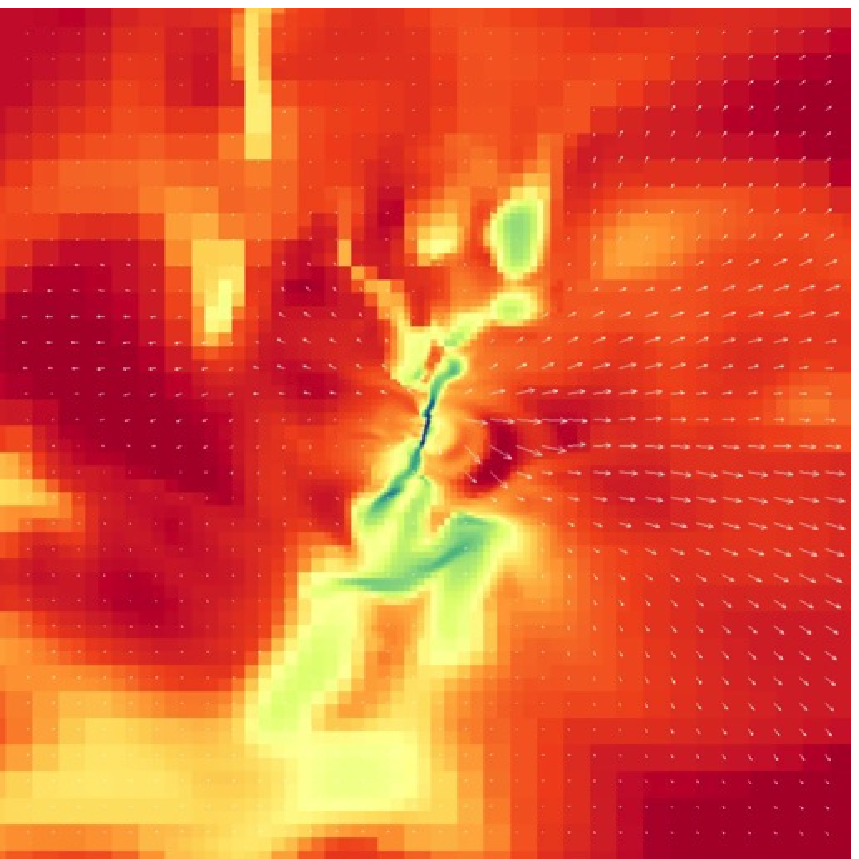}}}
  \caption{Slices of the gas density at $z=7.5$ for the \AGN~run where the energy is released through thermal bursts (top panel), or through collimated jets (bottom panel). Arrows indicate the velocity vectors in the plane of the image, and their lengths are proportional to the velocity amplitudes. The image sizes are 15~kpc. In both cases, an expanding wind develops from the central disc.}
    \label{fig:outflow}
\end{figure}

Our standard implementation of AGN feedback deposits energy in isotropic thermal bursts when BHs accrete at high rates ($\chi\ge0.01$) and collimated jets at low accretion rates ($\chi < 0.01$).
Both the thermal inputs and the jets deposit their energy (mass and momentum for the jet) in a small region that is approximated by a sphere of radius $\Delta x=15$~pc for the quasar mode, and a cylinder of radius $\Delta x$ and height $2\, \Delta x$ for the radio mode.
The size of the region of the energy input is typically two orders of magnitude smaller than the characteristic radius of the galactic disc (2 kpc at $z=6$).
In the radio mode, the momentum injected in each cell within the cylinder is parallel to the axis defined by the angular momentum vector of the gas surrounding the BH (maximum collimation with no opening angle).
As shown in Fig.~\ref{fig:sfr}, the BH accretes gas above $\chi\ge0.01$ all the way down to redshift $z=6$, thus energy is released through thermal pulses that should propagate isotropically in a homogeneous medium.
However, the BH is surrounded by a thin disc of dense star-forming gas.  As a result the outflow produced by the AGN in its quasar mode is not perfectly isotropic but exhibits a bipolar outflow with a large opening angle (Fig.~\ref{fig:outflow}).
It is possible that in reality winds are much more collimated, and that our predictions for the impact of AGN feedback on the galaxy and its surrounding gas are overestimated because the wind has a cross-section that is too large.
To test this, we run a simulation where AGN feedback is only allowed to release its energy in a collimated jet mode (the so-called radio mode).
The result is illustrated in the bottom panel of Fig.~\ref{fig:outflow}. The large-scale outflow produced by this jet mode is very similar to the isotropic thermal wind: a bipolar expanding wind develops around the central galaxy.
The reason for the quick expansion of the radio AGN is that the jet is light.
As a consequence the pressure inside the jet pushes the cavity surrounding the jet and the wind expands as much as it can in all directions~\citep{meieretal91}
The multiphase structure of the interstellar medium also help to destabilize the jet propagation~\citep{sutherland&bicknell07, gaibleretal11}.

\section{Reducing  AGN feedback efficiency}
\label{appendix:efficiency}

\begin{figure}
  \centering{\resizebox*{!}{6.5cm}{\includegraphics{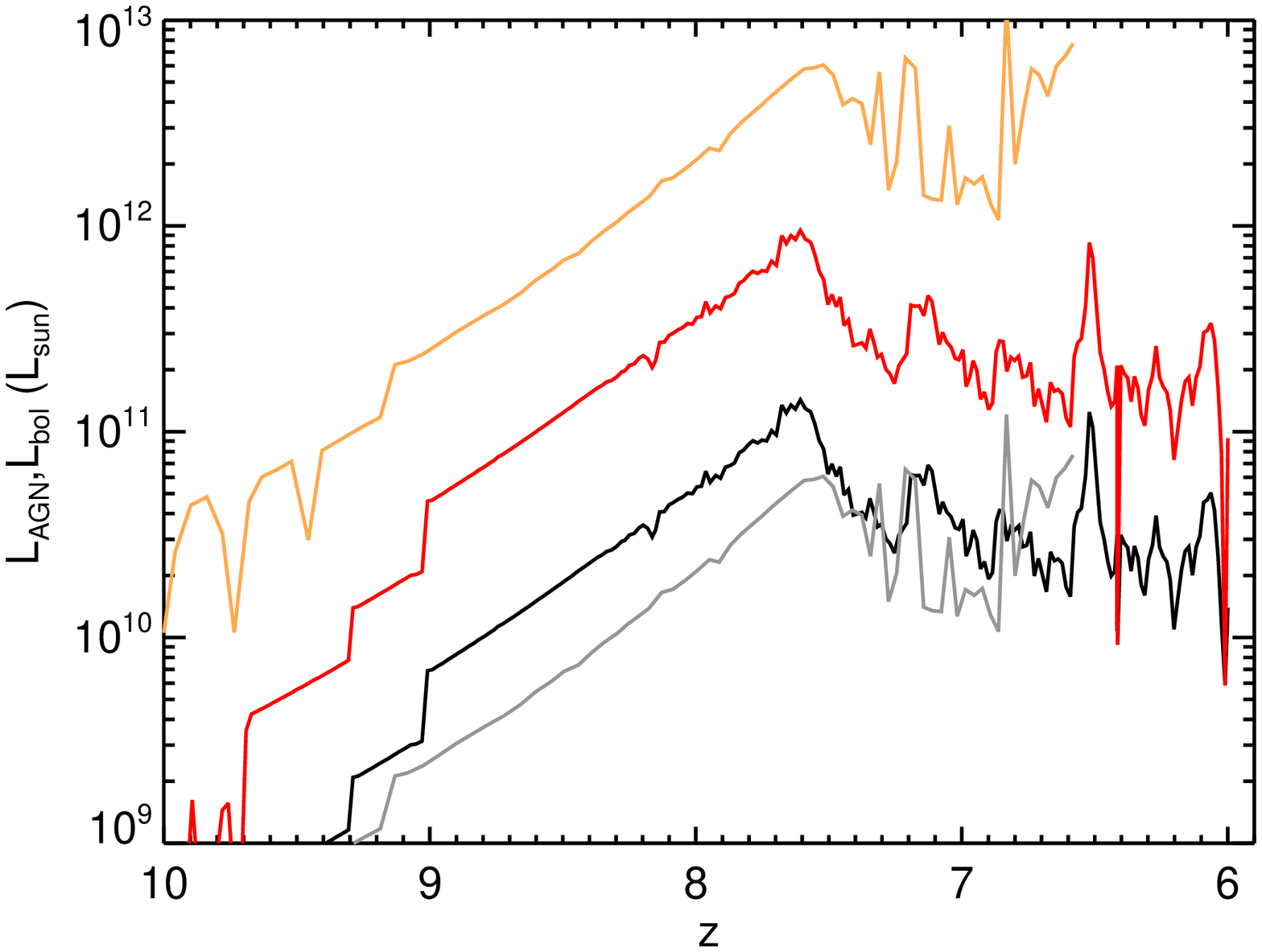}}}
  \caption{AGN feedback luminosity of the most massive BH for the canonical AGN model (black), the reduced efficiency reduced by 15 (grey), and their bolometric luminosities (red and orange respectively). The feedback luminosities are similar for the two AGN feedback models, but their bolometric luminosities are different by one order of magnitude. }
    \label{fig:lum2}
\end{figure}

\begin{figure}
  \centering{\resizebox*{!}{6.cm}{\includegraphics{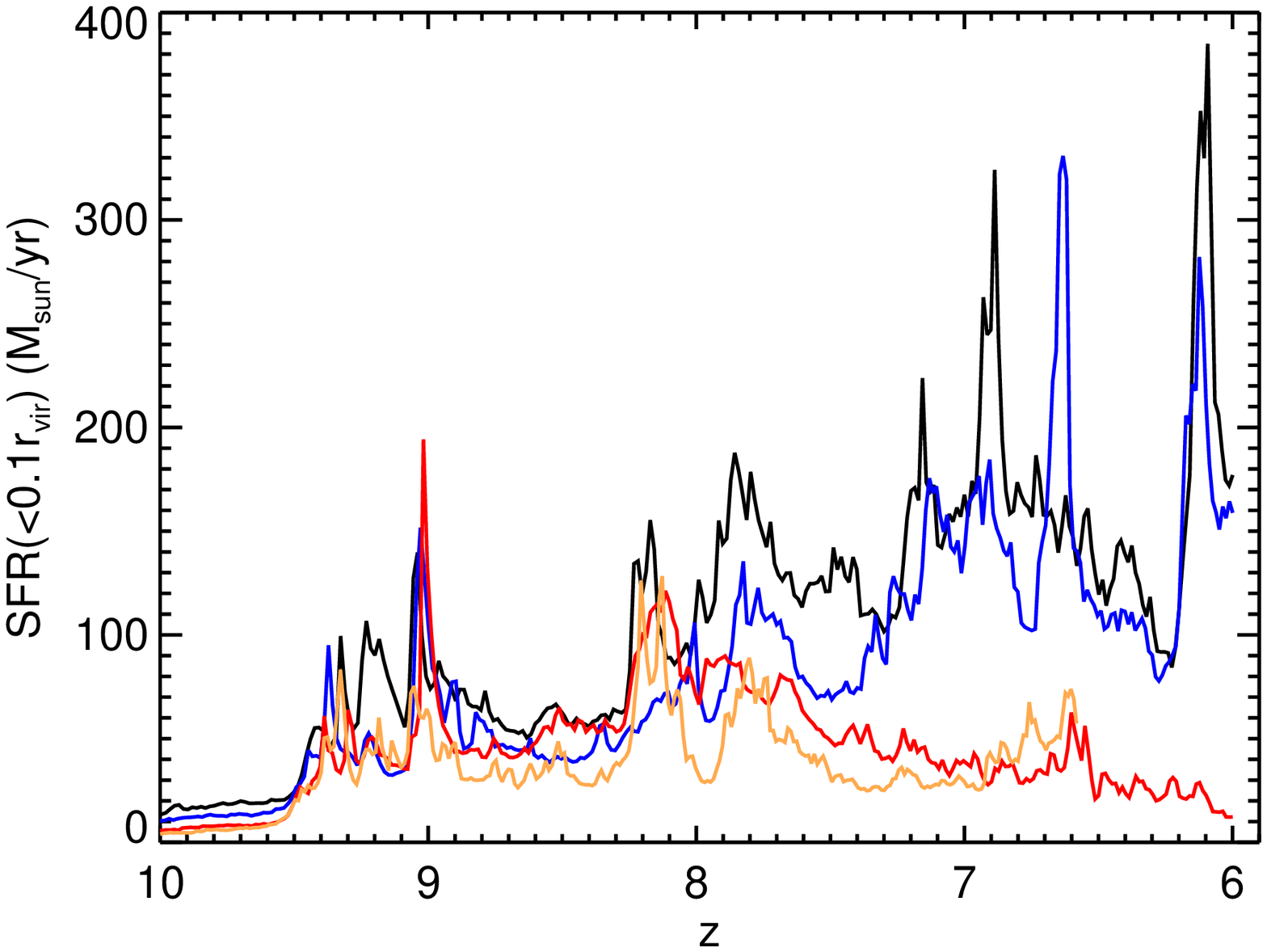}}}
  \caption{SFR history using the stars contained within $0.1\, r_{\rm vir}$ for the \nofeed~case (black), the \noAGN~case (blue), and for the \AGN~case with the canonical AGN feedback model (red), and the model with reduced efficiency (orange). The two AGN feedback models show similar SFRs with values reduced by a factor of 10 compared to \nofeed/\noAGN~cases.}
    \label{fig:sfr2}
\end{figure}

We also test how the growth of the BH and the impact of AGN feedback when the efficiency $\epsilon_{\rm f}$ of AGN energy coupling to the gas is reduced by a factor of 15.
In this case the central supermassive BH reaches a larger mass $M_{\rm BH, low}=5 \times 10^{8}\,\rm M_\odot$ at $z=6.7$ instead of $M_{\rm BH, high}=6\times 10^{7}\,\rm M_\odot$ for our canonical model at the same redshift.
Thus, the energy released by the two BHs are very similar (the energy released by the canonical model is 1.8 times the energy of model with reduced efficiency) because the energy required to unbind the dense cold gas surrounding the BH does not strongly vary with time.
BHs accrete close to, or at their Eddington rate before blowing out the gas and self-regulating their growth.
The characteristic growth time of a BH accreting at the Eddington limit is $t_{\rm Edd}=45$~Myr for a radiative efficiency of $\epsilon_{\rm r}=0.1$.
Thus, two BHs accreting gas at the Eddington limit with AGN feedback efficiencies $\epsilon_{\rm f}$ that are one order of magnitude different will reach the self-regulated state with a time delay of the order of $t_{\rm Edd}$ (see Fig.~\ref{fig:lum2}).
As the AGN feedback luminosities are comparable in both runs, the bolometric luminosity is naturally one order of magnitude larger in the simulation with reduced efficiency, and reaches values up to $10^{13}\, \rm L_\odot$ compatible with the bolometric luminosities observed for the brightest quasars at $z\sim6$~\citep[$10^{12}-10^{14}\, \rm L_\odot$,][]{willottetal10}.
As the energy released by the central BHs is comparable, the impact on the galaxy star formation is also similar (see Fig.~\ref{fig:sfr2}).
The two models of AGN feedback significantly reduce the SFR in the central object with levels of star formation that are comparable.
Hence the main results of this paper do not depend on the details of the feedback implementation and is only weakly dependent on the feedback efficiency $\epsilon_{\rm f}$.

\section{Resolution tests}
\label{appendix:resolution}

We added some other simulations where the minimum allowed cell size is degraded down to $\Delta x=135$~pc (above the nominal resolution used in the paper of $\Delta x=15$~pc) to test the influence of resolution on the impact of AGN feedback.
One of these simulations is the equivalent of the \nofeed~case (without SN and no AGN feedback), and another one is the equivalent of the \AGN~case (with SN and AGN feedback) described in section~\ref{section:numerics}. 
The change in resolution is accompanied by a modification in the threshold for star formation, which is now $\rho_0= 5\, \rm H\, cm^{-3}$.
The consequence is that the star forming gas is more supported by pressure due to a lower value of $\rho_0$ in the polytropic equation of state and produces a less clumpy medium with the star formation taking place more smoothly.
We added one last run by degrading both the DM resolution and spatial resolution to $M_{\rm res}=7\times 10^{6}\, h^{-1}\, \rm M_{\odot}$ and $\Delta x=135$~pc with the same physics as our canonical \AGN~case, and for which the threshold for star formation is $\rho_0= 5\, \rm H\, cm^{-3}$. 

\begin{figure}
  \centering{\resizebox*{!}{6.cm}{\includegraphics{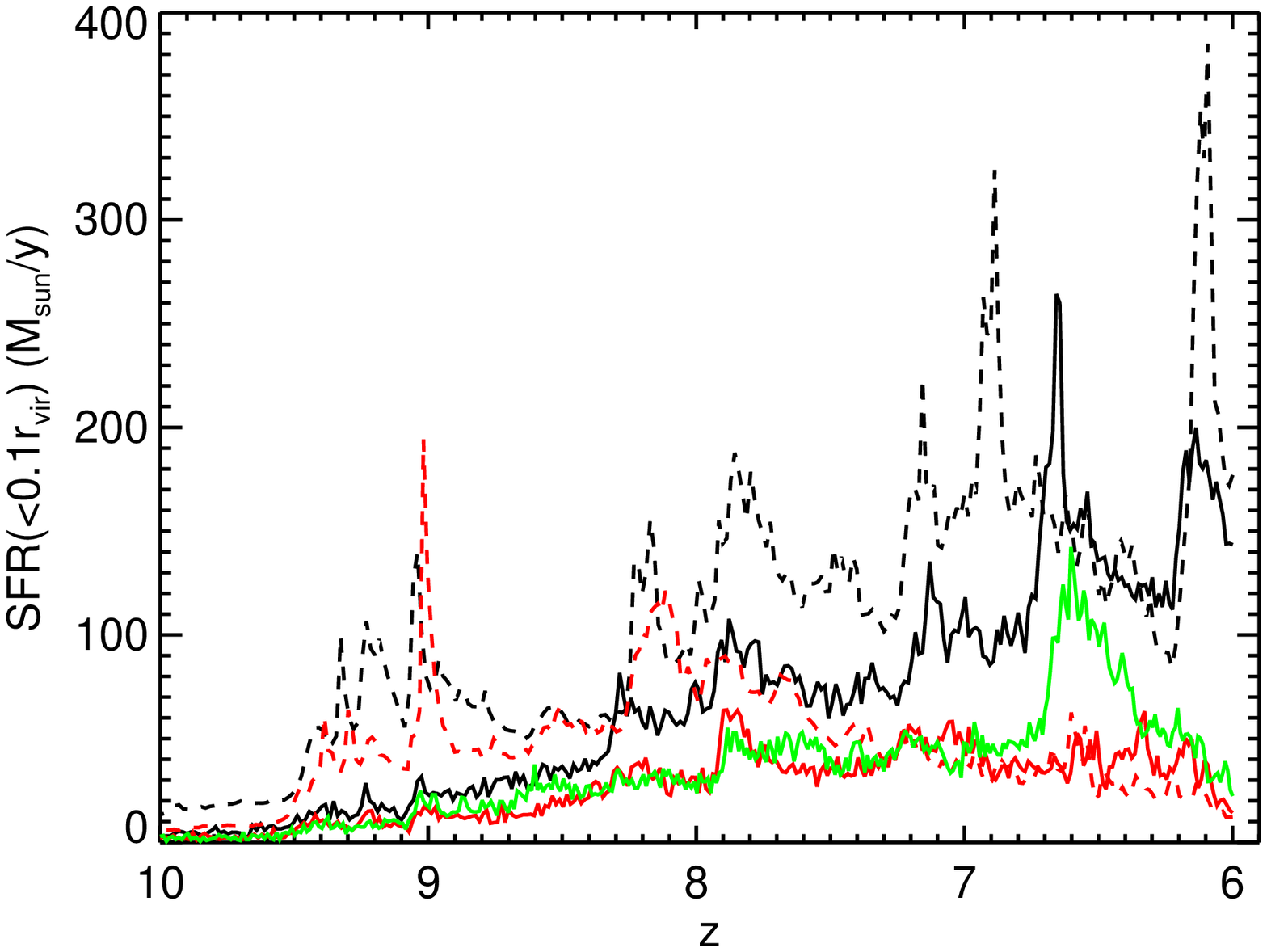}}}
  \caption{SFR history using the stars contained within $0.1\, r_{\rm vir}$ for the \nofeed~case with $\Delta x=15$~pc (black dashed), the \AGN~case with $\Delta x=15$~pc (red dashed), \nofeed~case with $\Delta x=135$~pc (black solid), the \AGN~case with $\Delta x=135$~pc (red solid), and the \AGN~case with $\Delta x=135$~pc and $M_{\rm res}=7\times 10^{6}\, h^{-1}\, \rm M_{\odot}$ (green). The SFRs with different resolution using the same physics have similar values at $z=6$, but they show a significant difference at higher redshift $z>7$.}
    \label{fig:sfrres}
\end{figure}

Fig.~\ref{fig:sfrres} shows the SFRs measured within $0.1\, r_{\rm vir}$ for these low resolution simulations and compared to the high resolution simulations.
The SFRs with the same physics but different resolutions have very similar values between $6<z<7$, which suggests that our results are well converged in this redshift range: the starvation of the gas on the central galaxy by AGN quenching has a comparable impact on the SFR at late times.
Above $z=7$, the SFRs are different from on resolution to another.
We interpret this as a result of both the force being sampled on coarser meshes, delaying the collapse of structures, and of a less clumpy and less dense ISM at low resolution, and as a consequence, the time-scales associated to star formation are longer (see~\citealp{teyssieretal10}).
The measured baryon fraction at $z=6$ and at $r_{\rm vir}$ is reduced by 30 per cent below the universal value for low resolution and high resolution \AGN~simulations. 
We also find that, at low resolution, the fraction of cold flows is reduced by a factor 2 at $0.1r_{\rm vir}$ due to the presence of AGN feedback compared to the lo-resolution \nofeed~case (at high resolution this reduction factor is 3, see Fig.~\ref{fig:mass4com}).
We can infer that the results found in this paper are robust to a dramatic change (one order of magnitude) in DM mass resolution and in minimum cell size.

\end{document}